\documentclass{elsart}
\usepackage{amssymb}
\usepackage{amsmath}
\usepackage[dvips]{graphicx}
\usepackage{color}
\journal{Journal of Computational Physics}
\begin{document}

\begin{frontmatter}
\title{Implementation of non-uniform FFT based Ewald summation in Dissipative Particle Dynamic method}
\author[]{Yong-Lei Wang$^{a,b}$},
\author[]{Aatto Laaksonen$^{a,}$\thanksref{Aatto}},
\author[]{Zhong-Yuan Lu$^{b,}$\thanksref{Lu}}
\thanks[Aatto]{Email: aatto.laaksonen@mmk.su.se}
\thanks[Lu]{Email: luzy@jlu.edu.cn}
\address{$^a$Department of Materials and Environmental Chemistry,\\Arrhenius Laboratory, \\Stockholm University, Stockholm, S-106 91, Sweden\\and\\ $^b$State Key Laboratory of Theoretical and Computational Chemistry, \\ Institute of Theoretical Chemistry, \\ Jilin University, Changchun 130023, China\\}

\begin{abstract}
The ENUF method, \emph{i.e.}, Ewald summation based on the Non-Uniform FFT technique (NFFT), is implemented in Dissipative Particle Dynamics (DPD) simulation scheme to fast and accurately calculate the electrostatic interactions at mesoscopic level. In a simple model electrolyte system, the suitable ENUF-DPD parameters, including the convergence parameter $\alpha$, the NFFT approximation parameter $p$, and the cut-offs for real and reciprocal space contributions, are carefully determined. With these optimized parameters, the ENUF-DPD method shows excellent efficiency and scales as $\mathcal{O}(N\log N)$. The ENUF-DPD method is further validated by investigating the effects of charge fraction of polyelectrolyte, ionic strength and counterion valency of added salts on polyelectrolyte conformations. The simulations in this paper, together with a separately published work of dendrimer-membrane complexes, show that the ENUF-DPD method is very robust and can be used to study charged complex systems at mesoscopic level.
\end{abstract}

\end{frontmatter}

\clearpage
\section{Introduction}
The Dissipative Particle Dynamics (DPD) simulation method~\cite{hoogerbrugge1992simulating, koelman1993dynamic} was originally developed to cover much longer length and time scales than in conventional atomistic Molecular Dynamics (MD) simulations using a mesoscopic description of the simulated system~\cite{espanol1995statistical} due to soft forces acting between large particles made of clusters of atoms. Since its introduction about two decades ago, several improvements and generalizations~\cite{groot1997dissipative, espanol1997dissipative, flekkoy1999molecular, pagonabarraga2001dissipative, warren2003vapor, espanol2003smoothed, nikunen2003would, maiti2004bead, jakobsen2005constant, travis2007new} have been proposed, making DPD method one of the mostly used coarse-grained approach in soft matter simulations. Examples of such studies are microphase separation of block copolymers~\cite{groot1998dynamic, qian2005computer}, polymeric surfactants in solution~\cite{ryjkina2002molecular, prinsen2002mesoscale}, colloidal suspensions~\cite{whittle2010dynamic}, and the structural and rheological behavior of biological membranes~\cite{shillcock2005tension, de2009effect}. For many of these complex systems mentioned above, the electrostatic interactions play a vital role behind the key phenomena and dynamical processes. The inclusion of electrostatic interactions in DPD simulations is important to capture the long-range interactions at mesoscopic level of material description for many systems such as the conformational properties of polyelectrolyte brushes~\cite{ibergay2009electrostatic, ibergay2010mesoscale, yan2009influence} and the formation of membrane-DNA complexes in biological systems~\cite{gao2009effects, gao2010communications}.

\par
However, when electrostatic interactions are incorporated in DPD method, the mesoscopic DPD particles carrying opposite charges tend to form artificial aggregates due to the soft nature of the conservative interactions (forces) in DPD simulations. In order to avoid a collapse of oppositely charged particles onto each other, Groot~\cite{groot2003electrostatic} smeared out the local point charge into grids around each DPD particle. He adopted a variant of particle-particle particle-mesh (PPPM) approach, which was originally introduced for systems with electrostatic heterogeneities~\cite{beckers1998iterative}, to treat separately the near and far field interactions between charge distributions. As the charge density distributions in the simulated system are affected by hydrodynamic flow~\cite{pagonabarraga2010recent}, this method provides a natural coupling between electrostatics and fluid motion. Simulation results~\cite{ibergay2009electrostatic, yan2009influence, gao2010communications, groot2003electrostatic} demonstrated that this method is reasonably efficient in capturing important features of electrostatic interactions at mesoscopic level.

\par
In an alternative approach, Gonz\'alez-Melchor~\emph{et al.}~\cite{gonzalez2006electrostatic} adopted traditional Ewald summation method and Slater-type charge density distribution in DPD simulations. This method allows a standard approach to calculate electrostatic energy and force of charge density distributions. Although the inclusion of charge density distribution does not directly increase the computational cost in the Ewald summation itself, this method becomes computationally more demanding than the one adopted by Groot~\cite{groot2003electrostatic}.

\par
In order to improve the efficiency and accuracy in treating electrostatic interactions, we suggest here an alternative approach that allows fast calculation of electrostatic energy and force in DPD simulations. The ENUF method, an abbreviation for the Ewald summation using Non-Uniform FFT (Fast Fourier Transform) (NFFT) technique, was recently suggested in our group~\cite{hedman2006ewald, hedman2006thesis} and showed excellent computational efficiency in atomistic MD simulations. In our current work, we implement the ENUF in the DPD method. The ENUF-DPD method is initially applied on simple electrolyte system as a typical model case to optimize the ENUF-DPD parameters and investigate corresponding computational complexity. With suitable parameters, we adopt the ENUF-DPD method to study the dependence of polyelectrolyte conformations on ionic strength and counterion valency of added salts for illustration.

\par
This paper is organized as follows: Sec.~\ref{sec:dpdmethod} contains a brief introduction to the DPD method. Detailed algorithm of the ENUF method is given in Sec.~\ref{sec:enuf}. Secs.~\ref{sec:enufdpd} and~\ref{sec:validation} describe the implementation of ENUF in DPD method and the exploration of suitable parameters for ENUF-DPD method in a simple model electrolyte system. In Sec.~\ref{sec:pe}, the ENUF-DPD method is further validated in studying polyelectrolyte conformations upon addition of salts with multivalent counterions. Finally, main concluding remarks are given in Sec.~\ref{sec:conclusion}.

\section{The DPD Method}\label{sec:dpdmethod}
The DPD method, originally introduced by Hoogerbrugge and Koelman in $1992$~\cite{hoogerbrugge1992simulating}, is a mesoscopic and particle-based simulation method based on a set of pairwise forces. One important conceptual difference between DPD and conventional MD is the use of coarse-graining (CG) procedure allowing a mapping of several atoms or molecules from the real atomistic system onto larger DPD particles. After scaling up the size of the system, the DPD method can capture the hydrodynamic behavior in very large complex systems up to microsecond range and beyond. Like in MD simulations, the time evolutions of DPD particles are governed by Newton's equations of motion
\begin{eqnarray} \label{eq01}
\frac{d\mathbf{r}_i}{dt}=\mathbf{v}_i\,,\qquad m_i\frac{d\mathbf{v}_i}{dt}=\mathbf{f}_i\,,
\end{eqnarray} where $\mathbf{r}_i$, $\mathbf{v}_i$ and $\mathbf{f}_i$ denote the coordinate, velocity, and the total force acting on particle $i$, respectively. The total force, between any pair of DPD particles $i$ and $j$, is normally composed of three different pairwise additive forces: the conservative force $\mathbf{F}_{ij}^{C}$, the dissipative force $\mathbf{F}_{ij}^D$, and the random force $\mathbf{F}_{ij}^R$,
\begin{eqnarray} \label{eq02}
\mathbf{f}_{ij}&=&\sum_{i\neq{j}}(\mathbf{F}_{ij}^{C}+\mathbf{F}_{ij}^{D}+\mathbf{F}_{ij}^{R})\,,
\end{eqnarray} with
\begin{eqnarray} \label{eq0305}
\mathbf{F}_{ij}^C&=&\alpha_{ij}\omega^C(r_{ij})\mathbf{\hat{r}}_{ij}\,,\\
\mathbf{F}_{ij}^D&=&-\gamma\omega^D(r_{ij})(\mathbf{v}_{ij}\cdot\mathbf{\hat{r}}_{ij})\mathbf{\hat{r}}_{ij}\,,\\
\mathbf{F}_{ij}^R&=&\sigma\omega^R(r_{ij})\theta_{ij}\mathbf{\hat{r}}_{ij}\,,
\end{eqnarray} where $\mathbf{r}_{ij}=\mathbf{r}_i-\mathbf{r}_j$, $r_{ij}=|\,\mathbf{r}_{ij}|$, $\mathbf{\hat{r}}_{ij}=\mathbf{r}_{ij}/r_{ij}$, and $\mathbf{v}_{ij}=\mathbf{v}_i-\mathbf{v}_j$. The parameters $\alpha_{ij}$, $\gamma$, and $\sigma$ determine the strength of the conservative, dissipative, and random forces, respectively. $\theta_{ij}$ is a randomly fluctuating variable, with zero mean and unit variance.

\par
The pairwise conservative force is written in terms of a weight function $\omega^C(r_{ij})$, where $\omega^C(r_{ij})=1-r_{ij}/R_c$ is chosen for $r_{ij}\leq R_c$ and $\omega^C(r_{ij})= 0$ for $r_{ij}>R_c$ such that the conservative force is soft and repulsive. The unit of length $R_c$ is related to the volume of DPD particles. In our simulations, we adopt the CG scheme~\cite{alsunaidi2004liquid} with $N_m=4$ and $\rho=4$, in which the former parameter means $4$ water molecules being coarse-grained into one DPD particle and the latter means there are $4$ DPD particles in the volume of $R_c^3$. With this particular scheme, the length unit $R_c$ is given as $R_c=3.107\sqrt[3]{\rho N_m}=7.829\textrm{\AA}$. The conservative interaction parameters between different types of DPD particles are determined by $\alpha_{ij}\approx \alpha_{ii}+2.05\chi$ with $\alpha_{ii}=78.67k_BT$, in which $\chi$ is the Flory-Huggins parameter between different types of DPD particles.

\par
Unlike the conservative force, two weight functions $\omega^D(r_{ij})$ and $\omega^R(r_{ij})$ for dissipative and random forces, respectively, are coupled together to form a thermostat. According to Espa\~nol and Warren~\cite{espanol1995statistical}, the relationship between two functions is described as
\begin{eqnarray} \label{eq06}
\left\{\begin{array}{l}
\omega^D\left(r\right)=\left[\omega^R\left(r\right)\right]^2  \\
\sigma^2=2\gamma k_BT
\end{array}\right.\,.
\end{eqnarray} This precise relationship between dissipative and random forces is determined by the fluctuation-dissipation theorem. We adopt a simple choice of $\omega^D(r)$ due to Groot and Warren~\cite{groot1997dissipative}
\begin{eqnarray} \label{eq07}
\omega^D\left(r\right)=\left[\omega^R\left(r\right)\right]^2=\left\{\begin{array}{ll}
 \left(1-r/R_c\right)^2&\left(r\leq R_c\right) \\
0&\left(r>R_c\right)
\end{array}\right.\,.
\end{eqnarray}

\par
For polymer and surfactant molecules, the intramolecular interactions between bonded particles are described by harmonic springs
\begin{eqnarray}
\mathbf{F}_i^S = - \sum_j K^S (r_{ij}-r_{eq})\mathbf{\hat{r}}_{ij}\,,
\end{eqnarray} where $K^S$ is the spring constant and $r_{eq}$ is the equilibrium bond length.

In traditional DPD method, charge density distributions are usually adopted instead of point charges to avoid the divergence of electrostatic interactions at $r=0$. In our implementation, a Slater-type charge density distribution with the form of
\begin{eqnarray} \label{eq26}
\rho_e(r)&=&\frac{q}{\pi\lambda_e^3}e^{\frac{-2r}{\lambda_e}}
\end{eqnarray} is adopted, in which $\lambda_e$ is the decay length of charge $q$. The integration of Eq.~\ref{eq26} over the whole space gives the total charge $q$.

\par
A modified version of velocity-Verlet algorithm~\cite{groot1997dissipative} is used to integrate the equations of motion. For easy numerical handling, we choose the cut-off radius, the particle mass, and $k_BT$ as the units of the simulating system, \emph{i.e.}, $R_c=m=k_BT=1$. As a consequence, the unit of time $\tau$ is expressed as $\tau=R_{c}\sqrt{m/k_BT}=1$. All the related parameters used in our DPD simulations are listed in Table~\ref{table:dpd-parameter}.

\section{The ENUF method}\label{sec:enuf}
\subsection{The Ewald summation method}
Consider a system composed of $N$ charged particles, each one carrying the partial charge $q_i$ at position $\textbf{r}_i$ in a cubic cell with the volume $V=L^3$. Overall charge neutrality is assumed in the simulations. For simplicity, only the simple charge-charge interaction is considered, interactions between dipoles and multipoles are omitted in our current scheme. Charges interact with each other according to the Coulomb's law, and the total electrostatic energy can be written as
\begin{eqnarray} \label{eq15}
\mathbf{U}^E(\textbf{r}^N)&=&\frac{1}{4\pi\epsilon_0\epsilon_r}\sum_{\textbf{n}}^{\dag}\sum_{i}\sum_{j>i} \frac{q_iq_j}{|\textbf{r}_{ij}+\textbf{n}L|}\,,
\end{eqnarray} where $\textbf{n}=(n_x,n_y,n_z)$, and $n_x$, $n_y$, and $n_z$ are integer numbers. The sum over $\textbf{n}$ takes into account the periodic images, and the \dag{} symbol indicates that the self-interaction terms are omitted when $\textbf{n}=0$. The variables $\epsilon_0$ and $\epsilon_r$ are the permittivity of vacuum and the dielectric constant of water at room temperature, respectively.

\par
In the Ewald summation method, the electrostatic energy, as shown in Eq.~\ref{eq15}, is decomposed into real space and reciprocal space contributions~\cite{frenkel1996understanding}. With such decomposition, both real and reciprocal space contributions are short-range and Eq.~\ref{eq15} can be rewritten as
\begin{eqnarray} \label{eq16}
\mathbf{U}^E(\textbf{r}^N)&=&\frac{1}{4\pi\epsilon_0\epsilon_r}\bigg\{\sum_{\textbf{n}}^{\dag}\sum_{j>i}\frac{q_iq_j}{\textbf{r}_{ij}+\textbf{n}L}\textrm{erfc}\big(\alpha
|\textbf{r}_{ij}+\textbf{n}L|\big) \nonumber \\
& & +\frac{2\pi}{V}\sum_{\textbf{n}\neq 0}\frac{e^{-|\textbf{n}|^{2}/4\alpha ^2}}{|\textbf{n}|^2}S(\textbf{n}) S(-\textbf{n}) -\frac{\alpha}{\sqrt{\pi}}\sum_j^N q_j^2\bigg\}\,,
\end{eqnarray} with
\begin{eqnarray}\label{eq17}
S(\textbf{n})=\sum_{i=1}^{N}q_ie^{-\imath \textbf{n}\cdot\textbf{r}_i} \quad \textrm{and} \quad \textbf{n}=\frac{2\pi}{L}(n_x,n_y,n_z)\,.
\end{eqnarray} The first, second, and last terms in the bracket on the right-hand side of Eq.~\ref{eq16} correspond to the electrostatic energies from real space, reciprocal space and self-interaction parts, respectively. $\alpha$ is the Ewald convergence parameter and determines the relative convergence rate between real and reciprocal space summations. $n$ is the magnitude of the reciprocal vector $\textbf{n}$. Choosing suitable parameters, the complexity of the Ewald summation method is reduced from $\mathcal{O}(N^{2})$ to $\mathcal{O}(N^{3/2})$ with considerably accuracy and efficiency~\cite{toukmaji1996ewald}.

\par
As the number of charged particles in the simulated system grows, it is convenient to combine the calculation of the short-range conservative force with the calculation of the real space summations of electrostatic interactions. The cut-off for conservative force calculations should be the same as the cut-off for real space electrostatic interactions. With such combination and suitable value of $\alpha$, the summation of real space electrostatic energy between two charged particles extends no longer than the cut-off distance, and can be expressed as
\begin{eqnarray}\label{eq31}
\mathbf{U}^{E,R}&=& \frac{1}{4\pi\epsilon_0\epsilon_r}\sum_i \sum_{j>i} \frac{q_iq_j}{r}\textrm{erfc}(\alpha r)\,.
\end{eqnarray} The real space electrostatic force on particle $i$ is the negative of the derivative of the potential energy $\mathbf{U}^{E,R}$ respect to its position $\textbf{r}_i$. A common form of the real space electrostatic force is described as
\begin{eqnarray}\label{eq32}
\mathbf{F}_i^{E,R}&=& -\nabla_i\mathbf{U}^{E,R} \nonumber\\
&=& \frac{1}{4\pi\epsilon_0\epsilon_r}\bigg\{\sum_{j\neq i}\frac{q_iq_j}{r^2} \textrm{erfc}(\alpha r)+\frac{2\alpha}{\sqrt{\pi}}e^{-\alpha^2 r^2}\bigg\}\,.
\end{eqnarray} Thus the real space electrostatic energy and force can be calculated together with conservative force, but then the computation of reciprocal space summations of electrostatic interactions becomes the more time-consuming part. The introduction of the FFT technique reduces the computational complexity of Ewald summation to $\mathcal{O}(N\log N)$ by treating reciprocal space summations with FFT technique~\cite{hedman2006ewald, toukmaji1996ewald, hockney1981computer, darden1993particle, luty1994comparison, essmann1995smooth}.

\subsection{Basic features of the FFT technique}
Initially for a finite number of given Fourier coefficients $\hat{\boldsymbol{f_k}}\in\boldsymbol{C}$ with $\boldsymbol{k}\in I_M$, we wish to evaluate the trigonometric polynomial
\begin{eqnarray}\label{eq08}
f(\boldsymbol{x})&=&\sum_{\boldsymbol{k}\in I_M}\hat{\boldsymbol{f_k}}e^{-2\pi \imath \boldsymbol{k}\cdot\boldsymbol{x}}
\end{eqnarray} at each of the $N$ given nonequispaced points
\begin{eqnarray*}
\boldsymbol{X}=\big\{ \boldsymbol{x}_j\in\boldsymbol{D}^d:j=0,1,\ldots,N-1\big\}
\end{eqnarray*}
which are randomly localized in $d$-dimensional domain
\begin{eqnarray*}
\boldsymbol{D}^d = \big\{\boldsymbol{x} = (x_t)_{t = 0,1,\ldots,d-1}:-\frac{1}{2}\leq x_t\leq\frac{1}{2}\big\}\,.
\end{eqnarray*}
The space of the $d$-variable function $f\in\boldsymbol{D}^d$ is restricted to the space of $d$-variable trigonometric polynomials $\left(e^{-2\pi \imath \boldsymbol{k}}:\boldsymbol{k}\in I_M \right)$ with degree $M_t$($t=0,1,\ldots,d-1$) in the $t$-th dimension. The possible frequencies $\boldsymbol{k}$ are collected in the multi index set $I_M$ with
\begin{eqnarray} \label{eq09}
I_M&=&\big\{\boldsymbol{k}=(k_t)_{t=0,1,\ldots,d-1}\in Z^d:-\frac{M_t}{2}\leq k_t\leq \frac{M_t}{2}\big\}\,.
\end{eqnarray} The dimension of the function space or the total number of points in the index set is $M_{\Pi}=\Pi_{t=0}^{d-1}M_t$.

\par
With these in prior definitions, the trigonometric polynomials for the $N$ given points can be described by
\begin{eqnarray} \label{eq10}
f_j=f(\boldsymbol{x}_j)=\sum_{\boldsymbol{k}\in I_M}\hat{\boldsymbol{f_k}}e^{-2\pi \imath \boldsymbol{k}\cdot\boldsymbol{x}_j}\qquad (j=0,1,\ldots,N-1)\,.
\end{eqnarray} Using the matrix-vector notation, all trigonometric polynomials can be rewritten as
\begin{eqnarray} \label{eq11}
\boldsymbol{f = A}\hat{\boldsymbol{f}}\,,
\end{eqnarray} where
\begin{eqnarray}\label{eq12}
\boldsymbol{f}&=&(f_j)_{j=0,1,\ldots,N-1}\,, \\ \nonumber
\boldsymbol{A}&=&(e^{-2\pi\imath\boldsymbol{k}\cdot\boldsymbol{x}_j})_{j=0,1,\ldots,N-1;\,\boldsymbol{k}\in I_M}\,,\\ \nonumber
\hat{\boldsymbol{f}}&=&(\hat{f}_{\boldsymbol{k}})_{\boldsymbol{k}\in I_M}\,.
\end{eqnarray}

\par
In the following implementations, the related matrix-vector products are the conjugated form
\begin{eqnarray} \label{eq13}
\boldsymbol{f}=\bar{\boldsymbol{A}}\hat{\boldsymbol{f}}\,, \qquad
\boldsymbol{f}_j=\sum_{\boldsymbol{k} \in I_M} \hat{\boldsymbol{f_k}} e^{2\pi \imath \boldsymbol{k} \cdot \boldsymbol{x}_j}\,,
\end{eqnarray} and the transposed form
\begin{eqnarray} \label{eq14}
\hat{\boldsymbol{f}}=\boldsymbol{A}^T\boldsymbol{f}\,, \qquad
\hat{\boldsymbol{f_k}}=\sum_{j=0}^{N-1}\boldsymbol{f}_je^{-2\pi \imath \boldsymbol{k}\cdot\boldsymbol{x}_j}\,,
\end{eqnarray} in which the matrix $\bar{\boldsymbol{A}}$ and $\boldsymbol{A}^T$ are the conjugated and transposed complex of the matrix $\boldsymbol{A}$, respectively. With given Fourier coefficients $\hat{\boldsymbol{f}}$, the Fourier samples $\boldsymbol{f}$ can be transformed with suitable FFT techniques in both directions.

\subsection{Detailed description of the ENUF method}
The ENUF method, which combines traditional Ewald summation method with the Non-Uniform fast Fourier transform technique, is a novel and fast method to calculate electrostatic interactions. The NFFT~\cite{press1989fast} is a generalization of the FFT technique~\cite{cooley1965algorithm}. The basic idea of NFFT is to combine the standard FFT and linear combinations of window functions which are well localized in both space and frequency domain. The controlled approximations using a cut-off in the frequency domain and a limited number of terms in the space domain result in an aliasing error and a truncation error, respectively. The aliasing error is controlled by the over-sampling factor $\sigma_s$, and the truncation error is controlled by the number of terms $p$. As described in Refs.~\cite{hedman2006ewald, hedman2006thesis}, NFFT only makes approximations on the reciprocal space part of Ewald summation. Hence in the following, we show the detailed procedure to calculate the reciprocal space summations of electrostatic energy and force with NFFT technique.

\par
Recasting the reciprocal space electrostatic energy in terms of Fourier components, the second term of Eq.~\ref{eq16} can be rewritten as
\begin{eqnarray} \label{eq18}
\mathbf{U}^{E,K}&=&\frac{1}{4\pi\epsilon_0\epsilon_r}\frac{2\pi}{V}\sum_{\textbf{n}\neq 0}\frac{e^{-|\textbf{n}|^{2}/4\alpha^2}}{|\textbf{n}|^2}S(\textbf{n})S(-\textbf{n})  \nonumber\\
&=&\frac{1}{4\pi\epsilon_0\epsilon_r}\frac{1}{2\pi L}\sum_{\textbf{n}\neq0} \frac{e^{-(\pi n)^2/(\alpha L)^2}}{n^2}S(\textbf{n})S(-\textbf{n})\,.
\end{eqnarray} For fixed vector $\textbf{n}$, the structure factor $S(\textbf{n})$ is just a complex number. With normalized locations, $\textbf{x}_j=\textbf{r}_j/L$, the structure factor can be expressed with
\begin{eqnarray} \label{eq19}
S(\textbf{n})&=&\sum_{j=1}^{N}q_je^{-2\pi \imath \textbf{n}\cdot\textbf{x}_j}\,.
\end{eqnarray} 
It should be noted that the structure factor in Eq.~\ref{eq19} and the transposed FFT form in Eq.~\ref{eq14} have similar structures. 
Suppose that $q_j$ is substituting $\boldsymbol{f}_j$, the structure factor $S(\textbf{n})$ is then a three-dimensional case of transposed FFT form. 
By viewing the structure factor $S(\textbf{n})$ as a trigonometric polynomial $\hat{\boldsymbol{f}}_{\textbf{n}}$, the reciprocal space electrostatic energy can be rewritten as
\begin{eqnarray} \label{eq20}
\mathbf{U}^{E,K}&=&\frac{1}{4\pi\epsilon_0\epsilon_r}\frac{1}{2\pi L}\sum_{\textbf{n}\neq 0} \frac{e^{-(\pi n)^2/(\alpha L)^2}}{n^2}|\hat{\boldsymbol{f}}_{\textbf{n}}|^2\,.
\end{eqnarray} 
The reciprocal space electrostatic energy is approximated by a linear combination of window functions sampled at nonequidistant $M_{\Pi}$ grids. 
These grids are then input to the transposed FFT, with which one can calculate each component of the structure factor $S(\textbf{n})$, and hence the reciprocal space summations of electrostatic energy.

\par
Alternatively, the reciprocal space electrostatic energy in Eq.~\ref{eq18} can be expressed by the real and imaginary parts of the structure factor $S(\textbf{n})$ as
\begin{eqnarray} \label{eq21}
\mathbf{U}^{E,K}&=&\frac{1}{4\pi\epsilon_0\epsilon_r}\frac{1}{2\pi L}\sum_{\textbf{n}\neq 0} \frac{e^{-(\pi n)^2/(\alpha L)^2}}{n^2}\bigg\{Re\big(S(\textbf{n})\big)^2 +Im\big(S(\textbf{n})\big)^2 \bigg\}\nonumber \\
&=&\frac{1}{4\pi\epsilon_0\epsilon_r}\frac{1}{2\pi L}\sum_{\textbf{n}\neq 0} \frac{e^{-(\pi n)^2/(\alpha L)^2}}{n^2}\bigg\{|\sum_iq_i\cos(\frac{2\pi}{L}\textbf{n} \cdot\textbf{r}_i)|^2\nonumber \\
&&{}+|\sum_iq_i\sin(\frac{2\pi}{L}\textbf{n}\cdot\textbf{r}_i)|^2\bigg\}\,.
\end{eqnarray} Similarly, the reciprocal space electrostatic force on particle $i$ is the negative derivative of the potential energy $\mathbf{U}^{E,K}$ respect to its position $\textbf{r}_i$ and is described as
\begin{eqnarray} \label{eq22}
\mathbf{F}_{i}^{E,K}&=&-\nabla_i\mathbf{U}^{E,K}\nonumber\\
&=&-\frac{1}{4\pi\epsilon_0\epsilon_r}\frac{1}{2\pi L}\sum_{\textbf{n}\neq0} \frac{e^{-(\pi n)^2/(\alpha L)^2}}{n^2}\left(\frac{4\pi q_i}{L}\textbf{n}\right) \nonumber \\
&&{}\bigg\{-\sin(\frac{2\pi}{L}\textbf{n}\cdot\textbf{r}_i)\sum_j q_j\cos(\frac{2\pi}{L}\textbf{n}\cdot\textbf{r}_j)\nonumber\\
&&{}+\cos(\frac{2\pi}{L}\textbf{n}\cdot\textbf{r}_i)\sum_j q_j\sin(\frac{2\pi}{L}\textbf{n}\cdot\textbf{r}_j)\bigg\} \nonumber \\
&=&\frac{1}{4\pi\epsilon_0\epsilon_r}\frac{2q_i}{L^2}\sum_{\textbf{n}\neq0} \textbf{n}\frac{e^{-(\pi n)^2/(\alpha L)^2}}{n^2}\bigg\{\sin(\frac{2\pi}{L}\textbf{n} \cdot\textbf{r}_i)Re\big(S(\textbf{n})\big)\nonumber \\
&&{}+\cos(\frac{2\pi}{L}\textbf{n}\cdot\textbf{r}_i)Im\big(S(\textbf{n})\big)\bigg\}\,.
\end{eqnarray} Since the structure factor $S(\textbf{n})$ is a complex number, the expression in the bracket of Eq.~\ref{eq22} can be written as the imaginary part of a product
\begin{eqnarray}\label{eq23}
\sin(\frac{2\pi}{L}\textbf{n}\cdot\textbf{r}_i)Re\big(S(\textbf{n})\big) + \cos(\frac{2\pi}{L}\textbf{n}\cdot\textbf{r}_i)Im\big(S(\textbf{n})\big) &=& Im\bigg\{e^{\frac{2\pi}{L}\imath\textbf{n}\cdot\textbf{r}_i}S(\textbf{n})\bigg\}\,.
\end{eqnarray} Following this expression, Eq.~\ref{eq22} can be expressed with
\begin{eqnarray} \label{eq24}
\mathbf{F}_{i}^{E,K}&=&\frac{1}{4\pi\epsilon_0\epsilon_r}\frac{2q_i}{L^2} \sum_{\textbf{n}\neq0}\textbf{n}\frac{e^{-(\pi n)^2/(\alpha L)^2}}{n^2} Im\bigg\{e^{\frac{2\pi}{L}\imath\textbf{n}\cdot\textbf{r}_i}S(\textbf{n})\bigg\} \nonumber\\
&=&\frac{1}{4\pi\epsilon_0\epsilon_r}\frac{2q_i}{L^2} Im\bigg\{\sum_{\textbf{n}\neq0}\textbf{n} \frac{e^{-(\pi n)^2/(\alpha L)^2}}{n^2}S(\textbf{n})e^{\frac{2\pi}{L}\imath\textbf{n}\cdot\textbf{r}_i}\bigg\}\nonumber\\
&=&\frac{1}{4\pi\epsilon_0\epsilon_r}\frac{2q_i}{L^2}Im\bigg\{\sum_{\textbf{n}\neq0} \hat{\textbf{g}}_{\textbf{n}}
e^{2\pi \imath\textbf{n}\cdot\textbf{x}_i}
\bigg\}\,,
\end{eqnarray} in which $\hat{\textbf{g}}_{\textbf{n}}=\textbf{n}\frac{e^{-(\pi n)^2/(\alpha L)^2}}{n^2}S(\textbf{n})$ with $\textbf{n}\neq0$. Again, Eq.~\ref{eq24} can be considered as a three-dimensional case of conjugated FFT form. Assuming $\textbf{n}\in I_M$ and $\hat{\textbf{g}}_0=0$, we can reformulate Eq.~\ref{eq24} into Fourier terms
\begin{eqnarray} \label{eq25}
\mathbf{F}_{i}^{E,K}&=&\frac{1}{4\pi\epsilon_0\epsilon_r}\frac{2q_i}{L^2} Im\bigg\{\sum_{\textbf{n}\in I_M}\hat{\textbf{g}}_{\textbf{n}}e^{2\pi \imath \textbf{n} \cdot\textbf{x}_i}\bigg\}\nonumber\\
&=&\frac{1}{4\pi\epsilon_0\epsilon_r}\frac{2q_i}{L^2}Im(\textbf{g}_i)\,.
\end{eqnarray} Thus, one can calculate the reciprocal space electrostatic force on particle $i$ using conjugated FFT based on the $S(\textbf{n})$ obtained from transposed FFT.

\par
From the equations shown above, it is clear that by using suitable NFFT techniques, the reciprocal space summations of both electrostatic energy and force can be calculated from ENUF method. The main features of ENUF method are listed in Table~\ref{table:algorithm}.

\section{The ENUF-DPD method}\label{sec:enufdpd}
In the traditional DPD formulation, the conservative force $\mathbf{F}_{ij}^C$ between interacting particles $i$ and $j$ is a short-range repulsive force, modeling the soft nature of neutral DPD particles. The electrostatic interactions between charged particles are long-range and conservative. In the ENUF-DPD simulations, the long-range electrostatic forces and short-range conservative forces are combined together to determine the thermodynamic behavior of the simulated systems~\cite{groot1997dissipative}.

\par
When the electrostatic interactions are included in DPD method, the main problem is that DPD particles with opposite charges show a tendency to collapse onto each other, forming artificial ionic clusters due to the soft nature of short-range repulsive interactions between DPD particles. In order to avoid this, the point charges at the center of DPD particles should be replaced by charge density distributions meshed around particles. Groot~\cite{groot2003electrostatic} firstly smeared out the local point charges around regular grids, and then adopted a variant of PPPM method to solve the near field and far field equations on grids instead of using FFT technique. Gonz\'alez-Melchor~\emph{et al.}~\cite{gonzalez2006electrostatic} directly adopted the Slater-type charge density distribution and traditional Ewald summation method to calculate the electrostatic interactions in DPD simulations. In our ENUF-DPD method, similar Slater-type charge density distribution and Gaussian type window functions in NFFT are adopted, respectively, to calculate real and reciprocal space contributions of electrostatic interactions.

\par
The detailed procedure of deducing electrostatic energy and force between two Slater-type charge density distributions are described as follows. The electrostatic potential field $\phi(r)$ generated by the Slater-type charge density distribution $\rho_e(r)$ can be obtained by solving the Poisson's equation
\begin{eqnarray} \label{eq:poisson_eq}
\nabla^2\phi(r) = -\frac{1}{\epsilon_0\epsilon_r}\rho_e(r)\,,
\end{eqnarray} In spherical coordinates, the Poisson's equation becomes
\begin{eqnarray} \label{eq:poisson_sph}
\frac{1}{r^2}\frac{\partial}{\partial r}\big(r^2\frac{\partial }{\partial r}\phi(r)\big) &=& -\frac{1}{\epsilon_0\epsilon_r}\rho_e(r) \,.
\end{eqnarray} For the Slater-type charge density distribution in Eq.~\ref{eq26}, we define parameter $c=-\frac{2}{\lambda_e}$ for the convenience of integration. Multiply $r^2$ at both sides of Eq.~\ref{eq:poisson_sph} and then integrate this equation, we can get
\begin{eqnarray} \label{eq:poisson_intea}
r^2\frac{\partial}{\partial r}\phi(r)&=&
-\frac{1}{\epsilon_0\epsilon_r}\frac{q}{\pi\lambda_e^3} 
\int_0^r(r')^2e^{cr'}dr \nonumber\\
&=&-\frac{1}{\epsilon_0\epsilon_r}\frac{q}{\pi\lambda_e^3}\Big[\big(\frac{r^2}{c}-\frac{2r}{c^2} +\frac{2}{c^3}\big)e^{cr}-\frac{2}{c^3}\Big] \,.
\end{eqnarray} By dividing $r^2$ at both sides of Eq.~\ref{eq:poisson_intea}, the potential field can be integrated analytically to give
\begin{eqnarray}\label{eq:potential_field}
\phi(r)&=&-\frac{1}{\epsilon_0\epsilon_r}\frac{q}{\pi\lambda_e^3}\int\Big[\big(\frac{1}{c}
-\frac{2}{c^2r}+\frac{2}{c^3r^2}\big)e^{cr}-\frac{2}{c^3r^2}\Big]dr \nonumber\\
&=&-\frac{1}{\epsilon_0\epsilon_r}\frac{q}{\pi\lambda_e^3}
\Big[\big(\int\frac{1}{c}e^{cr}dr\big)+\big(\int(\frac{2}{c^3r^2}-\frac{2}{c^2r})e^{cr}dr\big)
-\big(\int\frac{2}{c^3r^2}dr\big)\Big] \nonumber\\
&=&-\frac{1}{\epsilon_0\epsilon_r}\frac{q}{\pi\lambda_e^3}
\Big[\frac{1}{c^2}e^{cr}-\frac{2}{c^3r}e^{cr}+\frac{2}{c^3r}\Big] \,.
\end{eqnarray} With the definition of $c=-\frac{2}{\lambda_e}$, we can reformulate Eq.~\ref{eq:potential_field} as
\begin{eqnarray}
\phi(r)&=&-\frac{1}{\epsilon_0\epsilon_r}\frac{q}{\pi\lambda_e^3}\big(\frac{\lambda_e^2}{4}e^{\frac{-2r}{\lambda_e}}+\frac{\lambda_e^3}{4r}e^{\frac{-2r}{\lambda_e}}-\frac{\lambda_e^3}{4r} \big) \nonumber\\
&=&\frac{1}{4\pi\epsilon_0\epsilon_r}\frac{q}{r}\big(1-(1+\frac{r}{\lambda_e})e^{\frac{-2r}{\lambda_e}}\big)\,.
\end{eqnarray}

\par
The electrostatic energy between interacting particles $i$ and $j$ is the product of the charge of particle $i$ and the potential field generated by particle $j$ at position $r_i$
\begin{eqnarray}\label{eq:coulomb_energy}
\mathbf{U}_{ij}^{E,DPD}(r_{ij})&=&q_i\phi_j(r_i) \nonumber\\
&=&\frac{1}{4\pi\epsilon_0\epsilon_r}\frac{q_iq_j}{r_{ij}}
\big(1-(1+\frac{r_{ij}}{\lambda_e})e^{\frac{-2r_{ij}}{\lambda_e}}\big)\,.
\end{eqnarray} By defining dimensionless parameters $r^*=r/R_c$ as the reduced center-to-center distance between two charged DPD particles and $\beta=R_c/\lambda_e$, respectively, the reduced electrostatic energy between two Slater-type charge density distributions is given by
\begin{eqnarray}\label{eq27}
\mathbf{U}_{ij}^{E,DPD}(r_{ij}^*)&=&\frac{1}{4\pi\epsilon_0\epsilon_r}
\frac{q_iq_j}{R_cr_{ij}^*}\bigg\{1-\big(1+\beta r_{ij}^*\big)e^{-2\beta r_{ij}^*}\bigg\}\,.
\end{eqnarray}

\par
The electrostatic force on charged particle $i$ is
\begin{eqnarray}\label{eq:electro_force}
\mathbf{F}_{i}^{E,DPD}(r_{ij})&=&-\nabla_i\mathbf{U}_{ij}^{E,DPD}(r_{ij})\nonumber\\
&=&-\frac{q_iq_j}{4\pi\epsilon_0\epsilon_r}\bigg\{\nabla_i\Big[\frac{1}{r_{ij}} \big(1-(1+\frac{r_{ij}}{\lambda_e})e^{\frac{-2r_{ij}}{\lambda_e}}\big)\Big]\bigg\} \nonumber \\
&=&\frac{1}{4\pi\epsilon_0\epsilon_r}\frac{q_iq_j}{(r_{ij})^2}\bigg\{1-e^{\frac{-2r_{ij}}{\lambda_e}}-\frac{2r_{ij}}{\lambda_e}e^{\frac{-2r_{ij}}{\lambda_e}}-\frac{2r_{ij}^2}{\lambda_e^2}e^{\frac{-2r_{ij}}{\lambda_e}}
\bigg\}\,.
\end{eqnarray} With two dimensionless parameters $\beta$ and $r^*$, the magnitude of the reduced electrostatic force is
\begin{eqnarray}\label{eq28}
\mathbf{F}_{i}^{E,DPD}(r_{ij}^*)
&=&\frac{1}{4\pi\epsilon_0\epsilon_r}\frac{q_iq_j}{(R_cr_{ij}^*)^2}
\bigg\{1-\Big(1+2\beta r_{ij}^*(1+\beta r_{ij}^*)\Big)e^{-2\beta r_{ij}^*}\bigg\}\,.
\end{eqnarray} Comparing Eq.~\ref{eq15} with Eq.~\ref{eq27}, we can find that the electrostatic energy between two charges is scaled with correction factor $B_1=1-(1+\beta r^*)e^{-2\beta r^*}$ when the Slater-type charge density distributions are introduced in DPD simulations. Similarly, the electrostatic force between two charged particles is scaled with correction factor $B_2=1-\Big(1+2\beta r^*(1+\beta r^*)\Big)e^{-2\beta r^*}$ in DPD simulations.

\par
In the limit of $r_{ij}^* \to 0$, the reduced electrostatic energy and force between two charge density distributions are described by $\lim\limits_{r_{ij}^* \to 0}\mathbf{U}_{ij}^{E,DPD}(r_{ij}^*) = \frac{1}{4\pi\epsilon_0\epsilon_r} \frac{q_iq_j}{R_c}\beta$ and $\lim\limits_{r_{ij}^*\to 0}\mathbf{F}_{ij}^{E,DPD}(r_{ij}^*) = 0$, respectively. It is clear that the adoption of Slater-type charge density distributions in DPD simulations removes the divergence of electrostatic interactions at $r_{ij}^* =0$, which means that both electrostatic energy and force between two charged particles are finite quantities. By matching the electrostatic interactions between two charge density distributions at $r_{ij}^*=0$ with previous work~\cite{groot2003electrostatic} gives $\beta=1.125$. From the relation of $\beta=R_c/\lambda_e$, we can get $\lambda_e = 6.954~\textrm{\AA}$, which is consistent with the electrostatic smearing radii used in Gonz\'alez-Melchor's work~\cite{gonzalez2006electrostatic}.

\par
Fig.~\ref{figure:fig1} shows the representation of the reduced electrostatic energy and corresponding force with respect to the distance between two charged DPD particles. Both the electrostatic energy and force are calculated using ENUF-DPD method and Ewald summation method with Slater-type charge density distributions and reference parameters. For comparison, we also include the standard Coulombic potential and corresponding force, both of which do diverge at $r=0$. The electrostatic energy and force calculated using ENUF-DPD method are almost the same as those calculated using Ewald summation method with reference parameters. The positions of the maximum value of the electrostatic energy in both methods are basically the same, but the maximum value calculated from ENUF-DPD method is slightly smaller than that from the standard Ewald summation method within an acceptable statistical error. Comparing with the standard Coulombic potential and corresponding force, we find that both ENUF-DPD and Ewald summation methods can give indistinguishable energy and force differences at $r\geq 3.0R_c$. Hence, the ENUF method can capture the essential character of electrostatic interactions, as well as the Ewald summation method, at mesoscopic level~\cite{gonzalez2006electrostatic}.

\par
Combining the electrostatic force $\mathbf{F}_{ij}^{E,DPD}$ and the soft repulsive force $\mathbf{F}_{ij}^C$ gives the total conservative force $\mathbf{F}_{ij}^{C*}$ between interacting particles $i$ and $j$ in DPD simulations. The total conservative force $\mathbf{F}_{ij}^{C*}$, together with dissipative force $\mathbf{F}_{ij}^D$ and random force $\mathbf{F}_{ij}^R$, as well as the intramolecular bonding force $\mathbf{F}_i^S$ for polymers and surfactants, act on DPD particles and evolve toward equilibrium conditions before taking statistical analysis. The details of the update scheme for ENUF-DPD method in single integration step are shown in Table~\ref{table:algorithm}.

\section{The choice of ENUF-DPD parameters}\label{sec:validation}
As the number of charged DPD particles in the simulated system grows, the calculation of the reciprocal space electrostatic interactions will become the most time-consuming part. Using the suitable parameters in ENUF-DPD method assures that the time to calculate the real space summations is approximately the same as the time to calculate the reciprocal space summations, thereby reducing the total computational time. Herein we try to explore the ENUF-DPD related parameters and get a set of suitable parameters for following applications.

\par
The implementation of ENUF-DPD method uses the Ewald convergence parameter $\alpha$, required accuracy $\delta (\ll 1)$, and two cut-offs ($r_c$ for real space and $n_c$ for reciprocal space). These parameters are correlated with each other with the following two conditions
\begin{align}\label{eq29}
&e^{-\pi^2|\textbf{n}|^2/(\alpha L)^2}\leq\delta \Longrightarrow n_c\geq \frac{\alpha L}{\pi}\sqrt{-\log(\delta)}\Longrightarrow n_c\propto L\propto N^{1/3}\,, \nonumber \\
&\textrm{erfc}(\alpha r_c)\approx e^{-\alpha^2r_c^2}\leq\delta\Longrightarrow r_c\approx \frac{\pi n_c}{\alpha^2L}\,.
\end{align} With required $\delta$, it is more convenient to pick a suitable value for $n_c$. Then one can determine $\alpha$ and $r_c$ directly from Eq.~\ref{eq29}. However, due to the fact that $n_c$ should be integer and $r_c$ should be a suitable value for the cell-link list update scheme in DPD simulations, we adopt another procedure to get suitable parameters.

\par
First, we choose suitable $\delta$. It has been demonstrated that $\delta = 5.0\times10^{-5}$ is enough to keep acceptable accuracy in ENUF method~\cite{hedman2006ewald}. In DPD simulations, due to the soft repulsive feature of the conservative force $\mathbf{F}^C$ in Eq.~\ref{eq0305}, we adopt $\delta = 1.0\times10^{-4}$ in our ENUF-DPD method.

\par
Secondly, we determine suitable $r_c$ and $\alpha$. Gonz\'alez-Melchor~\emph{et al.}~\cite{gonzalez2006electrostatic} adopted $1.08 R_c$ and $3.0R_c$, respectively, as electrostatic smearing radii and real space cut-off for the calculation of electrostatic interactions with Ewald summation method. In our ENUF-DPD method, as specified in Eqs.~\ref{eq27} and~\ref{eq28}, electrostatic energy $\mathbf{U}^{E,DPD}$ and force $\mathbf{F}^{E,DPD}$ are scaled with two correction factors, $B_1$ and $B_2$, respectively, both of which are $r$-dependent. This means that the reciprocal space summations of electrostatic energy $\mathbf{U}^{E,K,DPD}$ and force $\mathbf{F}^{E,K,DPD}$ are also scaled with corresponding correction factors. It should be noted that what we obtain from FFT is the total influences of other particles on particle $i$. It is difficult to differentiate individual contribution since each corresponding correction factor is related to the relative distance between interacting particles. But if we choose suitable $r_c$, beyond which two correction factors $B_1$ and $B_2$ approximate to $1.0$, the total reciprocal space summations of electrostatic energy $B_1\mathbf{U}^{E,K,DPD}$ and force $B_2\mathbf{F}^{E,K,DPD}$ can be approximately expressed by $\mathbf{U}^{E,K,DPD}$ and $\mathbf{F}^{E,K,DPD}$, respectively. Such approximation enables us to adopt directly the FFT results as reciprocal space summations. In Fig.~\ref{figure:fig2}, we plot two correction factors $B_1$ and $B_2$ with respect to the distance $r$. It is clearly shown that both $B_1$ and $B_2$ approximate to 1.0 when $r \ge 3.0 R_c$. Hence in our simulations, $r_c = 3.0 R_c$ is taken as the cut-off for real space summations of electrostatic interactions. With such adoption, both $B_1$ and $B_2$ are only applied on real space summations of electrostatic interactions within cut-off $r_c =3.0 R_c$.

\par
Since the Fourier-based Ewald methods utilize the FFT technique to evaluate the reciprocal part summations, it is more appropriate to choose suitable $\alpha$, with which we can minimize the total computational time in calculating electrostatic interactions. A large value of $\alpha$ means that a small value of $r_c$ is used for rapid convergence in real space summations, but the reciprocal space calculations will be the more time-consuming part and vice versa. The choice of $\alpha$ is system-dependent and related to the trade-offs between accuracy and computational speed. Based on Eq.~\ref{eq29} and above determined $r_c=3.0R_c$, we deduce that $\alpha\geq 0.12~\textrm{\AA}^{-1}$. Although the electrostatic energy is invariant to the choice of $\alpha$, the $\alpha$ value indeed affects the total time in calculating electrostatic interactions. In order to find a suitable value for $\alpha$, we carry out our first simulation to evaluate the Madelung constant ($M$) of a face-centered cubic (FCC) lattice. The crystal structure is composed of $4000$ charged particles, half of which are cations with net charge $+1$ and the other half are anions carrying net charge $-1$. All ions are located on regular lattice of FCC structure. The electrostatic energy for the FCC crystal structure is calculated within single step and then used to determine Madelung constant $M$. Simulation details are listed in Table~\ref{table:test-detail}. By comparing our calculated $M$ values with the theoretical value in Ref.~\cite{lu2003computer}, we find that for a wide range of $\alpha$ values the calculated $M$ coincides with literature value. The lowest acceptable value, $\alpha=0.20~\textrm{\AA}^{-1}$, is then used in the following simulations to minimize the computational effort.

\par
Finally, with Eq.~\ref{eq29} and above described parameters, we can choose suitable value for $n_c$. For NFFT technique, there are two other parameters, $\sigma_s$ and $p$, controlling the approximation errors. It has been shown that for a fixed over-sampling factor $\sigma_s>1$, the error decays exponentially with $p$~\cite{hedman2006ewald, dutt1993fast}. In atomistic MD simulations, it has been shown that $\sigma_s=2$ is adequate to take enough samples~\cite{hedman2006ewald, hedman2006thesis}. Hence $\sigma_s=2$ is used in our ENUF-DPD simulations to keep a good accuracy.

\par
Since ENUF approximates only the reciprocal space summations of the standard Ewald summation, it is reasonable to expect that both the ENUF and the Ewald summation methods should behave in the same way at mesoscopic level. We then perform the second set of simulations on bulk electrolyte system to determine parameters, $n_c$ and $p$. The volume for each simulation is $V=(10R_c)^3$ with the total number of charged DPD particles $N=4000$, in which $2000$ charged DPD particles represent cations with net charge $+1$ and the same number of DPD particles carrying net charge $-1$ that represent the anions. All simulations are equilibrated for $1\times 10^4$ steps and then another $2\times10^4$ steps to take statistical average for the following analysis. Detailed simulation information are listed in Table~\ref{table:test-detail}.

\par
In Fig.~\ref{figure:fig3}, we present the relative errors of electrostatic energies calculated using ENUF-DPD and Ewald summation methods with explored parameters $n_c$ and $p$, as well as Ewald summation method with reference parameters. It is clear that as $n_c$ increases, the relative errors for electrostatic energy from Ewald summation method converge to $1.0\times10^{-4}$, which is the acceptable accuracy we set at first. For the ENUF-DPD method, the relative electrostatic energy errors generally decrease with the increase of $p$. When $p=1$, the relative errors from EUNF-DPD method converge to $2\times 10^{-3}$ for large $n_c$, indicating that the ENUF-DPD method with $p=1$ is too crude due to the fact that $p$ is not adequate to provide enough sampling terms in simulations. By increasing $p$, the relative errors reduce rapidly with the increase of $n_c$. In Fig.~\ref{figure:fig4}(a), we show three cases of the electrostatic energy errors calculated from ENUF-DPD method with $p=2$ and different $n_c$ values, as well as those from the Ewald summation method with same simulation parameters. It is obvious that for parameters $n_c\geq7$, the electrostatic energy errors, calculated from both ENUF-DPD and Ewald summation methods, fluctuate within $1\times10^{-4}$. This means that the electrostatic energy errors from ENUF-DPD and Ewald summation method with $p=2$ and $n_c=7$ are expected to be practically negligible in comparison with those calculated from traditional Ewald summation method with reference parameters, which gives the most precise electrostatic energies. Similar tendencies are also observed in the electrostatic force errors calculated from ENUF-DPD and Ewald summation methods with the same set of parameters, as shown in Fig.~\ref{figure:fig4}(b). In Fig.~\ref{figure:fig4}(c) we show the maximum errors in electrostatic forces. It is clear that as $p=2$, the increase of $n_c$ will enable ENUF-DPD method to be more and more similar to Ewald summation method. With larger $p$ values, such as $p=3$, we could further increase the accuracy of electrostatic interactions in ENUF-DPD method, but at the same time the total computational time in treating electrostatic interactions increases. By compromising the accuracy and computational speed in the ENUF-DPD method, we adopt $p=2$ and $n_c=7$ in following simulations.

\par
Now we perform the third set of simulations to study the structural properties of the electrolyte solution using Ewald summation method with reference parameters, ENUF-DPD and Ewald summation methods with above determined parameters, respectively. The system consists of $N=4000$ DPD particles in a simulation box of volume $V=(10R_c)^3$. The solvents mimicking water at room temperature are represented by $3736$ neutral DPD particles, and $132$ particles representing ions with net charge $+1$ and the same number of particles with net charge $-1$ representing counterions are also included in the simulations. Mapping these quantities to real units, the simulated system corresponds to dilute electrolyte solution with a salt concentration about $0.6$ M, which is consistent with the simulation condition in Refs.~\cite{groot2003electrostatic, gonzalez2006electrostatic}. Detailed simulation information can be found in Table~\ref{table:test-detail}.

\par
The structural properties of neutral solvents and charged particles are determined by the radial distribution functions (RDFs). The RDFs of neutral solvent-solvent, equal sign ions, and unequal sign ions are calculated from traditional Ewald summation with reference parameters, the Ewald summation and ENUF-DPD method with above determined parameters, as shown in Fig.~\ref{figure:fig5}. It is clear that the RDFs for the same pair particles calculated from three methods show similar tendencies. A general observation is that there is no ionic cluster formation at distance close to $r=0$. Furthermore, we also find that the RDFs between charged particles satisfy $g_{+-}(r)g_{++/--}(r)=g^2_{00}(r)$, where $+$, $-$, and $0$ correspond to positive charged, negative charged, and neutral DPD particles in simulations. This relationship between three RDFs implies that the structures between charged particles are related to the effective electrostatic interparticle potentials, as observed in Ref.~\cite{groot2003electrostatic}.

\par
Another important property of the ENUF-DPD method is its complexity in treating electrostatic interactions. Theoretically, the complexity for treating electrostatic interactions with FFT related technique is $\mathcal{O}(N\log N)$~\cite{toukmaji1996ewald}. For NFFT, the computational complexity is $\mathcal{O}(M_{\Pi}\log M_{\Pi} + \log(N/\delta))$, where $M_{\Pi}$ is the total number of points in the index set, and $\delta$ is the desired accuracy and also a function of $p$ for fixed over-sampling factor $\sigma_s$ in NFFT~\cite{press1989fast, dutt1993fast}. Combining the definition of $M_{\Pi}$ and the relationship in Eq.~\ref{eq29}, we can get $M_{\Pi}\propto n_c^3\propto N$, hence the theoretical complexity of ENUF-DPD method is $\mathcal{O}(N\log N+\log(N/\delta))$.

\par
The scaling behavior of the ENUF-DPD method, as well as the Ewald summation method, including its traditional version and the one with suitable parameters we have explored above, are estimated by varying the number of charged DPD particles in simulations. In each simulation, the total number of DPD particles is $N=32000$ with the volume $V=(20R_c)^3$. Initially, all DPD particles are neutral, and corresponding computational time is taken as the benchmark for DPD simulations without electrostatic interactions. Then we perform a number of different simulations increasing the number of charged particles up to $24000$ while keeping the simulation system neutral and the total number of DPD particles fixed. The detailed simulation information can be found in Table~\ref{table:test-detail}. Fig.~\ref{figure:fig6} shows the averaged run time per $10^3$ steps of DPD simulation as function of the number of charged particles. The averaged execution time per $10^3$ steps is the net time in calculating the electrostatic energy and force in each DPD simulation. It reveals that the original Ewald summation method with reference parameters could generate accurate electrostatic energy and force, but its computational complexity is $\mathcal{O}(N^2)$. With our above determined suitable parameters, the scaling behavior of Ewald summation method is reduced to $\mathcal{O}(N^{3/2})$ in general. Although the parameters for Ewald summation method in our simulations are a little different from the parameters used by Gonz\'alez-Melchor~\emph{et al.}~\cite{gonzalez2006electrostatic}, the computational complexities are described by similar scaling behavior $\mathcal{O}(N^{3/2})$, which has also been verified by Ibergay~\emph{et al.}~\cite{ibergay2009electrostatic}. The ENUF-DPD method scales as $\mathcal{O}(N\log N)$, which is in line with the scaling behavior of PPPM method adopted in Groot's work~\cite{ibergay2009electrostatic, groot2003electrostatic}.

\par
The ENUF-DPD method with above explored parameters shows an excellent computational efficiency and good accuracy in treating electrostatic interactions. In the current study of electrolyte solution and the range of the number of ion pairs investigated here, the ENUF-DPD method performs clearly much faster than the Ewald summation method, and shows similar $\mathcal{O}(N\log N)$ scaling behavior as PPPM method~\cite{ibergay2009electrostatic, groot2003electrostatic} at mesoscopic level.

\section{Conformation of polyelectrolyte in solution}\label{sec:pe}
The above implemented ENUF-DPD method has all capabilities of ordinary DPD, but includes applications where electrostatic interactions are essential but previously inaccessible. One key example is the polyelectrolyte conformation. Electrostatic interactions between charged particles on polyelectrolyte lead to rich conformation of polyelectrolyte qualitatively different from those of uncharged polymers~\cite{dobrynin2005theory, jusufi2009colloquium}. In this section, we use the ENUF-DPD method to study the charge fraction of polyelectrolyte on the conformational behavior of single polyelectrolyte molecule.

\par
Nine charge fractions, defined by $f=N_q / N_p$, in which $N_q$ and $N_p$ are, respectively, the number of charged particles and the total number of particles on polyelectrolyte, are considered. In our simulations, $N_p=48$ is used. Although this generic polyelectrolyte model is much smaller than the real ones, the essential physical effects are still captured~\cite{dobrynin2005theory, de1995precipitation, liu2003polyelectrolyte, kirwan2004imaging}. Going from a neutral polymer chain to a fully ionized polyelectrolyte, $f$ takes $0.0$, $0.125$, $0.25$, $0.375$, $0.5$, $0.625$, $0.75$, $0.875$ and $1.0$, which correspond to $N_q=0$, $6$, $12$, $18$, $24$, $30$, $36$, $42$ and $48$ charged particles on polyelectrolyte, respectively. Each charged particle on polyelectrolyte is characterized by net charge $-1$. Counterions carrying net charge $+1$ are added to preserve charge neutrality of simulation systems. The equilibrium bond distance between bonded particles is set to $r_{eq}=0.7R_c$ and the spring constant is taken as $K^S=64.0k_BT$. The conservative interaction parameters between different types of DPD particles are obtained through $\alpha_{ij} \approx \alpha_{ii}+2.05\chi$ with $\alpha_{ii}=78.67k_BT$. The $\chi$ parameter between polyelectrolyte and solvents is set to $0.87$, which is rescaled from Refs.~\cite{groot2003electrostatic, gonzalez2006electrostatic}.

\par
All simulations are performed in a simulation cell with volume $V = (30R_c)^3$. The total density is fixed at $\rho=4$, and hence the total number of particles in the system is $N=108000$ in all cases. All simulations are equilibrated in $5 \times 10^4$ time steps, and then $2.5 \times 10^5$ time step simulations are further performed to collect statistical data. For a fully ionized polyelectrolyte, additional simulations with larger volume $V = (40R_c)^3$ are also performed. No differences beyond statistical uncertainties are found between the two sets of simulations. In the following discussion, all simulation results are calculated from systems with volume $V = (30R_c)^3$. Simulation details are listed in Table~\ref{table:test-detail}.

\par
The averaged radius of gyration $<$$R_g$$>$ of polyelectrolyte, as function of corresponding charge fraction $f$, are shown in Fig.~\ref{figure:fig7}. Experimentally, it is well known that a neutral polymer in solution has the smallest radius of gyration~\cite{stigter1995theory, roiter2005afm, liao2006counterion}. When partial groups on polyelectrolyte are ionized, such as weakly charged polyelectrolyte in solution with adjustable pH values, the $<$$R_g$$>$ of polyelectrolyte increases with increasing degree of ionization of polyelectrolyte~\cite{roiter2005afm}. For fully ionized polyelectrolyte, the value of $<$$R_g$$>$ is approximately $1.31$ times larger than that of neutral polymer, which is consistent with molecular simulation results~\cite{gonzalez2006electrostatic} and experimental observations~\cite{roiter2005afm}. Typical conformations of polyelectrolyte with charge fraction $f=0.0$, $0.25$, $0.5$, $0.75$ and $1.0$ are shown in Fig.~\ref{figure:fig8}. These simulation results are qualitatively consistent with experimental observations~\cite{kirwan2004imaging, roiter2005afm} and theoretical predictions~\cite{stigter1995theory, liao2006counterion} for weakly charged polyelectrolyte.

\par
With the increase of charge fraction $f$ on polyelectrolyte, the RDFs between charged particles on polyelectrolyte and monovalent counterions are also enhanced, as shown in Fig.~\ref{figure:fig9}(a). In Fig.~\ref{figure:fig9}(b), we show the intramolecular pair correlation functions between charged particles of polyelectrolyte. A quantitative measure of the structure is found by analyzing the intramolecular pair correlation functions. For neutral polymer and polyelectrolyte with various charge fractions, the intramolecular correlations in initial zone $r/R_c < 1$ are dominated by particle-particle repulsions. In the regime $r/R_c >1$, two striking tendencies are shown in Fig.~\ref{figure:fig9}(b). For polyelectrolyte with small charge fraction, \emph{i.e.}, $f < 0.25$, we observe a small scaling-like domain and then followed by a terminal correlation range. In contrast, polyelectrolyte with large charge faction shows a scaling behavior over the entire range. The slope of the fitting in Fig.~\ref{figure:fig9}(b) is $-1.92$, which is in good agreement with Groot's results~\cite{groot2003electrostatic}.

\par
When salts are added into solution, both ionic strength and valency of multivalent counterions of added salts can severely influence the conformational properties of polyelectrolyte due to the strong correlations between multivalent counterions and polyelectrolyte. This behavior is usually specified as the overcharging phenomenon that occurs in many biological and synthetic polyelectrolytes~\cite{sanders2005structure}. Herein, the ENUF-DPD method is further adopted to investigate the effects of ionic strength and counterion valency of added salts on the conformation behavior of fully ionized polyelectrolyte.

\par
The number of multivalent counterions of added salts, $N_c$, is determined by $\theta=qN_c/N_p$, where $\theta$ is the ratio between the total charge of multivalent counterions of added salts and that of polyelectrolyte, and $q$ is the valency of added salts. In our simulations, various $\theta$ values, together with $q = 1$, $2$, and $3$, are selected to consider the dependence of fully ionized polyelectrolyte conformation on the ionic strength and valency of multivalent counterions of added salts. All simulations are equilibrated in $5 \times 10^4$ time steps, and $2.5 \times 10^5$ time step simulations are further performed to collect statistical data. Detailed simulation information are listed in Table~\ref{table:test-detail}.

\par
Fig.~\ref{figure:fig10} shows the dependence of $<$$R_g$$>$ on $\theta$ and valency of multivalent counterions of added salts. In the absence of added salts, polyelectrolyte adopts an extended conformation, owing to the electrostatic repulsions between charged particles of polyelectrolyte. Upon addition of salts, these repulsions are screened and hence $<$$R_g$$>$ decreases. For monovalent counterions ($q = 1$), $<$$R_g$$>$ gradually decreases. By contrast, stronger decreases in $<$$R_g$$>$ are observed in the cases of added salts with divalent ($q = 2$) and trivalent counterions ($q = 3$), occurring at considerably lower ionic strength. This reflects the conformational collapse of polyelectrolyte in solution with multivalent counterions, which has been observed in experiments~\cite{liu2003polyelectrolyte, roiter2010single} and predicted by coarse-grained molecular dynamic simulations~\cite{hsiao2006salt}.

\par
The smallest value of $<$$R_g$$>$ occurs near $c_Z$, \emph{i.e.}, the (Z:$1$) salt concentration at which the total charge of the Z-valent counterions of added salts neutralizes the bare polyelectrolyte. At the same time, polyelectrolyte shows compact conformation. Accordingly, this compact state occurs at a salt concentration that decreases with the increase of corresponding counterion valency, which is consistent with the two-state model~\cite{solis2000collapse}. In addition, multivalent counterions with higher valency are strongly correlated with polyelectrolyte, which can be specified by corresponding pair correlation functions. The RDFs among polyelectrolyte, monovalent counterions of polyelectrolyte, and multivalent counterions of added salts at $\theta=1.0$, are calculated and shown in Fig.~\ref{figure:fig11}. In solutions, monovalent counterions of polyelectrolyte and multivalent counterions of added salts show different condensation abilities on polyelectrolyte. For ($1$:$1$) salt, two kinds of counterions show similar tendencies due to the same amount of net charge on them. With the increase of counterion valency of added salts, counterions with different valency show competition in condensating the polyelectrolyte. The peak of RDF between trivalent counterions and polyelectrolyte is much higher than that of other counterions, implying the strong condensation between trivalent counterions and polyelectrolyte. The strong condensation induced by electrostatic correlations decreases the osmotic pressure, and hence leads to the collapse of polyelectrolyte~\cite{mei2006collapse}. Typical conformations of polyelectrolyte, as well as added salt with the counterion valency $q=1$, $2$, and $3$, are shown in Fig.~\ref{figure:fig12}.

\par
A striking effect occurs once the salt concentration is increased beyond $c_Z$. Polyelectrolyte starts to swell, in close analogy with the redissolution observed for multichain aggregates~\cite{de1995precipitation}. Comparing with conformation of polyelectrolyte in ($1$:$1$) salt, which exhibits a slow, monotonic decrease of $<$$R_g$$>$ with the increase of salt concentrations, the slight swelling behavior of polyelectrolyte in the presence of multivalent counterions emphasizes the important role of counterion valency~\cite{liu2003polyelectrolyte, roiter2010single}.

\par
Concerning the effects of ionic strength and valency of added salts on polyelectrolyte conformation, the decay of correlations between charge particles can be specified by the Debye screening length. The addition of salts with multivalent counterions leads to a short Debye screening length~\cite{yan2009dissipative}, demonstrating that the electrostatic interactions between charge particles separated larger than specific distance become screened and hence are no longer long-range. It is very likely that a finite cut-off for electrostatic interactions, or a screened interaction potential between charge particles, can be used in handling electrostatic interactions in these cases. This topic is beyond the scope of present work, deserving a special attention and consideration in future.

\par
The other detailed application of the ENUF-DPD method is the specific binding structures of dendrimers on amphiphilic membranes~\cite{yonglei}. We construct mutually consistent coarse-grained models for dendrimers and lipid molecules, which can properly describe the conformation of charged dendrimers and the surface tension of amphiphilic membranes, respectively. Systematic simulations are performed and simulation results reveal that the permeability of dendrimers across membranes is enhanced upon increasing dendrimer sizes. The negative curvature of amphiphilic membrane formed in dendrimer-membrane complexes is related to dendrimer concentration. Higher dendrimer concentration together with the synergistic effect between charged dendrimers can also enhance the permeability of dendrimers across amphiphilic membranes. Detailed descriptions of this work are shown in Ref.~\cite{yonglei}.

\section{Conclusion}\label{sec:conclusion}
The ENUF method, which combines the traditional Ewald summation method with NFFT technique to calculate the electrostatic interactions in MD simulations, is incorporated in DPD method. The ENUF-DPD method is applied on simple model electrolyte systems to explore suitable parameters. With required accuracy parameter $\delta=1.0\times10^{-4}$ and cut-off $r_c=3.0R_c$ for real space summations of electrostatic interactions, we find that the Ewald convergence parameter $\alpha=0.20~\textrm{\AA}^{-1}$ can generate accurate Madelung constant for FCC lattice structure and keep considerable accuracy. Simulation results reveal that the ENUF-DPD method with approximation parameter $p=2$ in NFFT and cut-off $n_c=7$ for reciprocal space summations of electrostatic interactions can well describe the electrostatic energy and force, as well as the Ewald summation method does. The computational complexity of ENUF-DPD method is approximately described as $\mathcal{O}(N\log N)$, which shows remarkably better efficiency than the traditional Ewald summation method with acceptable accuracy in treating long-range electrostatic interactions between charged particles at mesoscopic level.

\par
The ENUF-DPD method is further validated by investigating the influence of charge fraction of polyelectrolyte on corresponding conformational properties. Meanwhile, the dependence of the conformations of fully ionized polyelectrolyte on ionic strength and valency of added salts are also studied. These applications, together with a separately published research work on the formation of dendrimer-membrane complexes, show that the ENUF-DPD method is very robust and can be used to study charged complex systems at mesoscopic level.

\section{Acknowledgments}
We gratefully acknowledge financial support from the Swedish Science Council (VR) and generous computing time allocation from SNIC. This work is subsidized by the National Basic Research Program of China ($973$ Program, $2012$CB$821500$), and supported by National Science Foundation of China ($21025416$, $50930001$).

\clearpage
\begin{table}[!htbp]
\begin{center}
\caption{Summary of the ENUF and DPD parameters used in this work.}\label{table:dpd-parameter}
\begin{tabular}{l@{\qquad}l@{\qquad}l}
\hline\hline
Parameter & Value & Meaning \\
\hline
\multicolumn{3}{c}{DPD parameters} \\
 $N_m$      &  $4$    & Number of water molecules represented by one DPD particle  \\
 $\rho$     &  $4$    & Number of DPD particles in the volume of $R_c^3$           \\
 $R_c$      &$7.829\textrm{\AA}$ & Range of soft repulsive interaction in DPD method \\
 $\alpha_{ii}$&$78.67$& Maximum repulsion parameter between identical DPD particles  \\
 $\chi_{ij}$ & $0.87$  & Flory-Huggins parameter between polyelectrolyte and solvents \\
  $K^S$     &   $64.0$   & Spring constant for bonded interactions of polyelectrolyte \\
  $r_{eq}$  &   $0.7$    & Equilibrium bond length of polyelectrolyte               \\
 $\gamma$   &   $6.74$   & Dissipation strength                                     \\
 $\sigma$   &   $3.67$   & Noise amplitude                                          \\
 $\lambda$  &  $0.65$    & Velocity prediction parameter in integration algorithm  \\
 $\delta_t$ &  $0.02\tau$& Integration time step                                    \\
 $m$        &  $1$       & Reduced particle mass                                    \\
 $k_BT$     &  $1$       &  Energy unit                                 \\
\multicolumn{3}{c}{ENUF parameters} \\
 $\delta$   &  $10^{-4}$ &  Accuracy of electrostatic interactions \\
 $  r_c $   &  $3.0R_c$  &Cut-off for real space summations of electrostatic calculation \\
 $\alpha$   &$0.20\textrm{\AA}^{-1}$ &  Ewald convergence parameter \\
 $\sigma_s$ & $2$        &  Over-sampling factor             \\
    $p$     & $2$        &  Approximation parameter in NFFT \\
 $  n_c $   & $7$   & Cut-off for reciprocal space summations of electrostatic calculation \\
 $\lambda_e$& $6.954\textrm{\AA}$&  Charge decay length \\
 $\epsilon_0 \epsilon_r$& $1$&  Dielectric constants \\
\hline\hline
\end{tabular}
\end{center}
\end{table}

\clearpage
\begin{table}[!htbp]
\begin{center}
\caption{The modified version of velocity-Verlet algorithm
for the ENUF-DPD method in single integration step.}\label{table:algorithm}
\begin{tabular}{l}
\hline\hline
(0) $\mathbf{v}_i^0 \leftarrow \mathbf{v}_i + \lambda \frac{1}{m} \left(\mathbf{F}_i^{C*} \Delta{t}+ \mathbf{F}_i^D \Delta{t} + \mathbf{F}_i^R \sqrt{\Delta{t}} \right)$\\
(1) $\mathbf{v}_i \leftarrow \mathbf{v}_i + \frac{1}{2}\frac{1}{m} \left(\mathbf{F}_i^{C*} \Delta{t} + \mathbf{F}_i^D\Delta{t} + \mathbf{F}_i^R \sqrt{\Delta{t}} \right)$\\
(2) $\mathbf{r}_i \leftarrow \mathbf{r}_i + \mathbf{v}_i \Delta{t}$\\
(3) Calculate $\mathbf{F}_i^C\{\mathbf{r}_j\}, \mathbf{F}_i^D\{\mathbf{r}_j,\mathbf{v}_j^0\}, \mathbf{F}_i^R\{\mathbf{r}_j\}$ and $\mathbf{F}_i^S\{\mathbf{r}_j\}$\\
(4) Calculate $\mathbf{F}_i^{E,DPD}$ and $\mathbf{U}^{E,DPD}$ \\
\qquad (i) Calculate $\mathbf{F}_i^{E,R}$ and $\mathbf{U}^{E,R}$ \\
\qquad (ii) Calculate $\mathbf{F}_i^{E,K}$ and $\mathbf{U}^{E,K}$ \\
\qquad\qquad $S(\textbf{n}) =\hat{\boldsymbol{f}}_{\textbf{n}} \leftarrow  \sum_{i=1}^{N}q_ie^{-2\pi\imath\textbf{n}\cdot\textbf{x}_i}$ (using transposed NFFT) \\
\qquad\qquad $\mathbf{U}^{E,K} \leftarrow \sum_{\textbf{n}\neq 0} f(n) S(\textbf{n})S(\textbf{-n})$ \\
\qquad\qquad $\hat{\textbf{g}}_{\textbf{n}} \leftarrow \textbf{n}\frac{e^{-(\pi n)^2/(\alpha L)^2}}{n^2}S(\textbf{n})$ \\
\qquad\qquad $\textbf{g}_i \leftarrow \sum_{\textbf{n}\in I_M}\hat{\textbf{g}}_{\textbf{n}}e^{2\pi \imath\textbf{n} \cdot\textbf{x}_i}$ (using conjugated NFFT) \\
\qquad\qquad $\mathbf{F}_{i}^{E,K} \leftarrow Im(\textbf{g}_i)$ \\
\qquad (iii) Calculate $\mathbf{U}^{E,SI}$ (Self-interaction energy)\\
\qquad (iv) $\mathbf{U}^{E,DPD} \leftarrow B_1 (\mathbf{U}^{E,R}+ \mathbf{U}^{E,K}  - \mathbf{U}^{E,SI})$ \\
\qquad  (v) $\mathbf{F}_i^{E,DPD} \leftarrow B_2 (\mathbf{F}_i^{E,R} + \mathbf{F}_i^{E,K})$ \\
(5) $\mathbf{F}_i^{C*} \leftarrow \mathbf{F}_i^C + \mathbf{F}_i^S + \mathbf{F}_i^{E,DPD}$ \\
(6) $\mathbf{v}_i \leftarrow \mathbf{v}_i + \frac{1}{2}\frac{1}{m}\left(\mathbf{F}_i^{C*}\Delta{t} + \mathbf{F}_i^D\Delta{t} + \mathbf{F}_i^R \sqrt{\Delta{t}}\right)$\\
\hline\hline
\end{tabular}
\end{center}
\end{table}

\clearpage
\begin{table}[!htbp]
\begin{center}
\caption{Simulation details in this work. The $1^{st}$ set of simulations is conducted to calculate Madelung constant $M$ and determine Ewald convergence parameter $\alpha$. All charged particles are generated on FCC lattice. The $2^{nd}$ set of simulations is carried out to determine two ENUF parameters, $p$ and $n_c$. The $3^{rd}$ and $4^{th}$ simulations are used to calculate the RDFs between neutral and charged DPD particles, and to investigate the scaling behavior of the ENUF-DPD method, respectively. The $5^{th}$ and $6^{th}$ simulations are performed to study the effects of charge fraction of polyelectrolyte, ionic strength and counterion valency of added salts on polyelectrolyte conformations, respectively. Representations: $N^T$, total number of DPD particles in simulation; $N^+$, number of cations; $N^-$, number of anions; $N^0$, number of neutral particles; $f$, charge fraction of polyelectrolyte; $N_p$, total number of particles on polyelectrolyte; $\theta$, charge ratio between multivalent counterions of added salts and that of polyelectrolyte; $q$, counterion valency of added salts.}\label{table:test-detail}
\begin{tabular}{l@{\quad}l@{\quad}l@{\quad}l@{\quad}l@{\quad}l@{\quad}l}
\hline\hline
    &$1^{st}$  & $2^{nd}$ & $3^{rd}$  &  $4^{th}$   & $5^{th}$ & $6^{th}$ \\
\hline
Volume   & $(10R_c)^3$ & $(10R_c)^3$ & $(10R_c)^3$  & $(20R_c)^3$ & $(30R_c)^3$ & $(30R_c)^3$\\
$N^T$    & $4000$      & $4000$      & $4000$       & $32000$   & $108000$  & $108000$  \\
$N^+$    & $2000$      & $2000$      & $132$        & $0\sim12000$  & $fN_p$  & $48+\frac{\theta N_p}{q}$  \\
$N^-$    & $2000$      & $2000$      & $132$        & $0\sim12000$  & $fN_p$ & $48+ \theta N_p$  \\
$N^0$    &  $0$        &  $0$        & $3736$       &$32000-2N^+$  &$108000-2fN_p$ &$107904-\theta N_p(\frac{1+q}{q})$\\
Steps    &  $1$        &$3\times10^4$&$2.1\times10^5$&$3\times10^4$  &$3\times10^5$&$3\times10^5$ \\
\hline\hline
\end{tabular}
\end{center}
\end{table}

\clearpage
\begin{figure}[!ht]
\begin{center}
\caption{Electrostatic potential and force between two charged DPD particles are calculated from ENUF and Ewald summation with reference parameters. For comparison, the standard Coulombic potential and force, both of which diverge at $r=0$, are also included. Both the potential and the force expressions are plotted for two equal sign charge density distributions.}\label{figure:fig1}
\end{center}
\end{figure}

\begin{figure}[!ht]
\begin{center}
\caption{Two correction factors, $B_1$ and $B_2$, are plotted with respect to the distance $r$. The dotted horizontal and vertical lines imply the value of $1.0$ and the cut-off $r_c = 3.0 R_c$ for real space summations of electrostatic interactions, respectively.}\label{figure:fig2}
\end{center}
\end{figure}

\begin{figure}[!ht]
\begin{center}
\caption{The relative errors of electrostatic energy calculated using the ENUF-DPD and Ewald summation with varied $p$ and $n_c$. The relative errors are defined as $|\frac{\mathbf{U}-\mathbf{U}^{ref}}{\mathbf{U}^{ref}}|$, where $\mathbf{U}^{ref}$ is the total electrostatic energy calculated from Ewald summation method with reference parameters, $\mathbf{U}$ is the total electrostatic energy calculated via either the ENUF-DPD or Ewald summation with corresponding explored parameters.}\label{figure:fig3}
\end{center}
\end{figure}

\begin{figure}[!ht]
\begin{center}
\caption{The errors in electrostatic energy ($\Delta\mathbf{U}$), force ($\Delta\mathbf{F}$) and corresponding maximum values ($\textrm{max} |\Delta\mathbf{F}|$) calculated using ENUF-DPD (red squares) and Ewald summation (green circles) with $p=2$ and various $n_c$, are compared with those calculated from traditional Ewald summation with reference parameters. (a)~$\Delta\mathbf{U}=\frac{(\mathbf{U}-\mathbf{U}_{ref})}{\mathbf{U}_{ref}}$, where $\mathbf{U}_{ref}$ is the total electrostatic energy calculated from Ewald summation with reference parameters, $\mathbf{U}$ is the total electrostatic energy calculated via either ENUF-DPD or Ewald summation with determined parameters. (b)~$\Delta\mathbf{F}=\frac{\mathbf{\bar{F}}^E-\mathbf{\bar{F}}^{E,ref}}{\mathbf{\bar{F}}^{E,ref}}$, where $\mathbf{\bar{F}}^{E,ref}$ is the averaged electrostatic force on DPD particles calculated from Ewald summation with reference parameters, $\mathbf{\bar{F}}^E$ is the average electrostatic force on charged particles calculated via either ENUF-DPD or Ewald summation with determined parameters. (c)~$\textrm{max} |\Delta\mathbf{F}| = \textrm{max} |\frac{\mathbf{F}_i^E-\mathbf{F}_i^{E,ref}}{\mathbf{F}_i^{E,ref}}|_{i=1,2,\cdots,N}$, where $\mathbf{F}_i^{E,ref}$ is the electrostatic force on particle $i$ calculated from Ewald summation with reference parameters, $\mathbf{F}_i^E$ is the electrostatic force on particle $i$ calculated via either ENUF-DPD or Ewald summation with determined parameters.}\label{figure:fig4}
\end{center}
\end{figure}

\begin{figure}[!ht]
\begin{center}
\caption{The pair correlation functions between different types of DPD particles calculated from Ewald summation with reference parameters, ENUF-DPD and Ewald summation with determined parameters.}\label{figure:fig5}
\end{center}
\end{figure}

\begin{figure}[!ht]
\begin{center}
\caption{The scaling behavior of ENUF-DPD method with determined parameters in our simulations. The averaged time per $10^3$ steps is calculated form DPD simulations carried out over $2\times10^4$ steps. For comparison, we also show the scaling behavior of Ewald summation method, including its traditional version with reference parameters and the one with parameters the same as those used in our ENUF-DPD method.}\label{figure:fig6}
\end{center}
\end{figure}

\begin{figure}[!ht]
\begin{center}
\caption{The averaged radius of gyration of polyelectrolyte as function of charge fraction $f$ going from neutral polymer chain to fully ionized polyelectrolyte.}\label{figure:fig7}
\end{center}
\end{figure}

\begin{figure}[!ht]
\begin{center}
\caption{Typical conformations of polyelectrolyte with different charge fraction $f$. Solvent particles are not shown for clarity. The red and cyan spheres in all images indicate the charged and neutral particles of polyelectrolyte, respectively. In addition, counterions are represented by yellow spheres. (a) $f=0.0$, (b) $f=0.25$, (c) $f=0.5$, (d) $f=0.75$, and (e) $f=1.0$.}\label{figure:fig8}
\end{center}
\end{figure}

\begin{figure}[!ht]
\begin{center}
\caption{(a)~The intermolecular RDFs between charged particles of polyelectrolyte and corresponding monovalent counterions. (b)~The intramolecular RDFs between charge particles of polyelectrolyte.}\label{figure:fig9}
\end{center}
\end{figure}

\begin{figure}[!ht]
\begin{center}
\caption{The dependence of $<$$R_g$$>$ of polyelectrolyte on $\theta$ and valency of multivalent counterions of added salts. The dash curve gives the $<$$R_g$$>$ of neutral polymer in solution without salts.}\label{figure:fig10}
\end{center}
\end{figure}

\begin{figure}[!ht]
\begin{center}
\caption{The pair correlation functions between polyelectrolyte and counterions, including monovalent counterions of polyelectrolyte and multivalent counterions of added salts.}\label{figure:fig11}
\end{center}
\end{figure}

\begin{figure}[!ht]
\begin{center}
\caption{Typical conformations of polyelectrolyte with added salts at $\theta=1$. Solvent particles are not shown for clarity. The red and yellow spheres indicate the charged particles of polyelectrolyte and corresponding counterions, respectively. Monovalent, divalent and trivalent counterions (cations) of added salts are represented by purple, green and blue spheres, respectively. All anions of added salts are presented by magenta spheres. (a) ($1$:$1$) salt, (b) ($2$:$1$) salt, and (c) ($3$:$1$) salt.}\label{figure:fig12}
\end{center}
\end{figure}

\clearpage
\setcounter{figure}{0}
\newpage
\begin{figure}[!ht]
\begin{center}
\includegraphics[angle=0,width=1.0\linewidth]{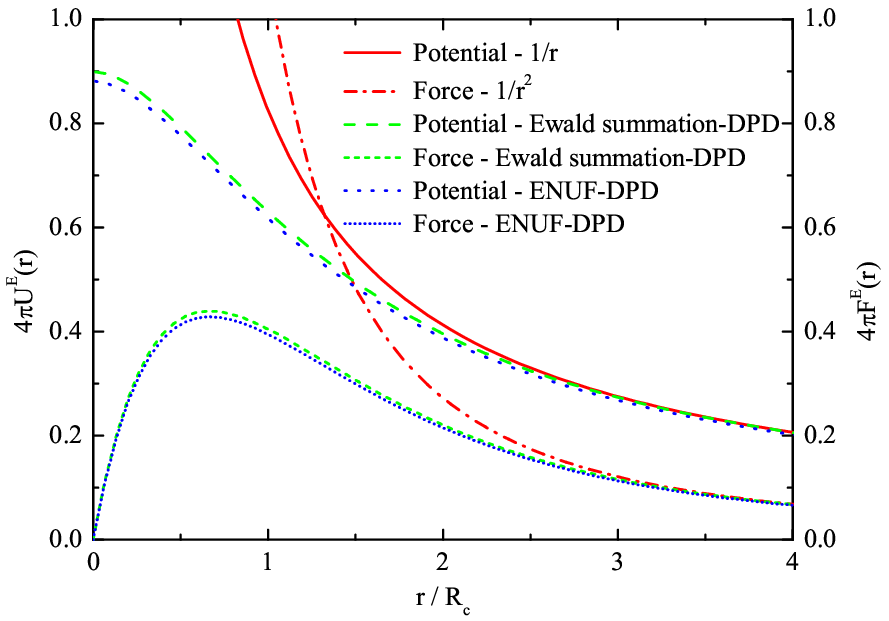}
\caption{Wang \emph{et al.} FIGURE 1}
\end{center}
\end{figure}

\clearpage
\begin{figure}[!ht]
\begin{center}
\includegraphics[angle=0,width=1.0\linewidth]{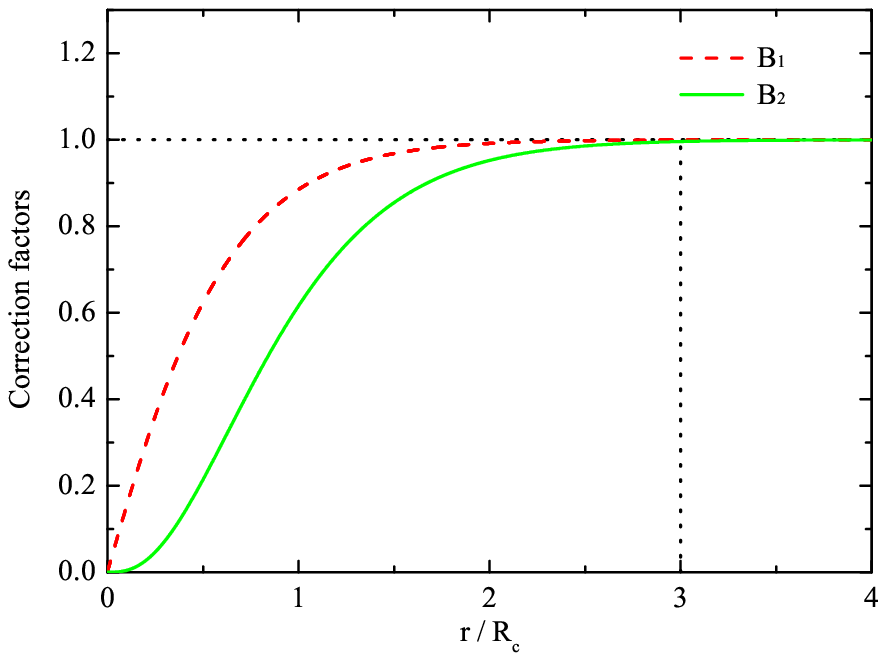}
\caption{Wang \emph{et al.} FIGURE 2}
\end{center}
\end{figure}

\clearpage
\begin{figure}[!ht]
\begin{center}
\includegraphics[angle=0,width=1.0\linewidth]{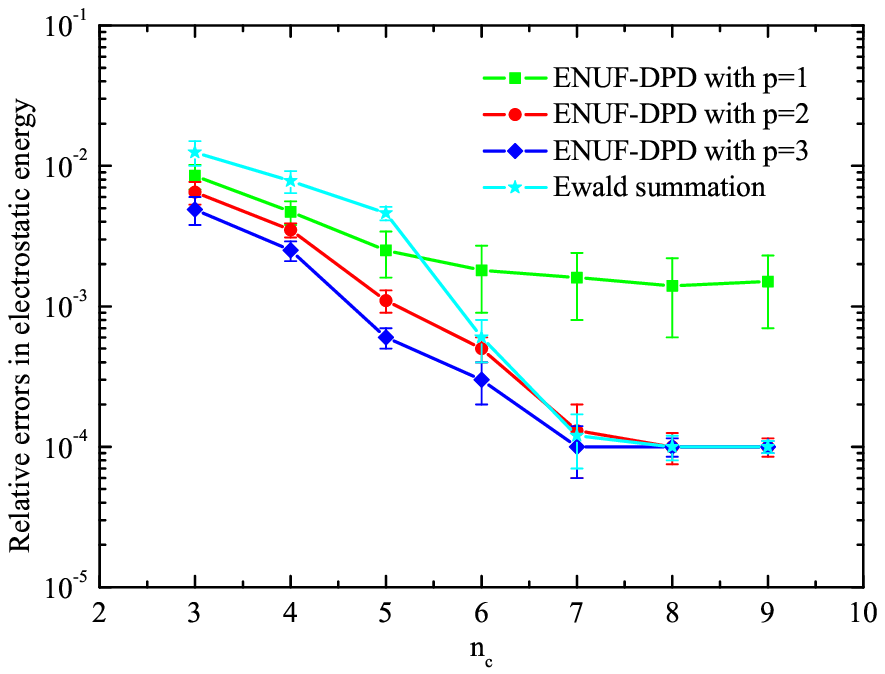}
\caption{Wang \emph{et al.} FIGURE 3}
\end{center}
\end{figure}

\clearpage
\begin{figure}[!ht]
\begin{center}
\includegraphics[angle=0,width=1.0\linewidth]{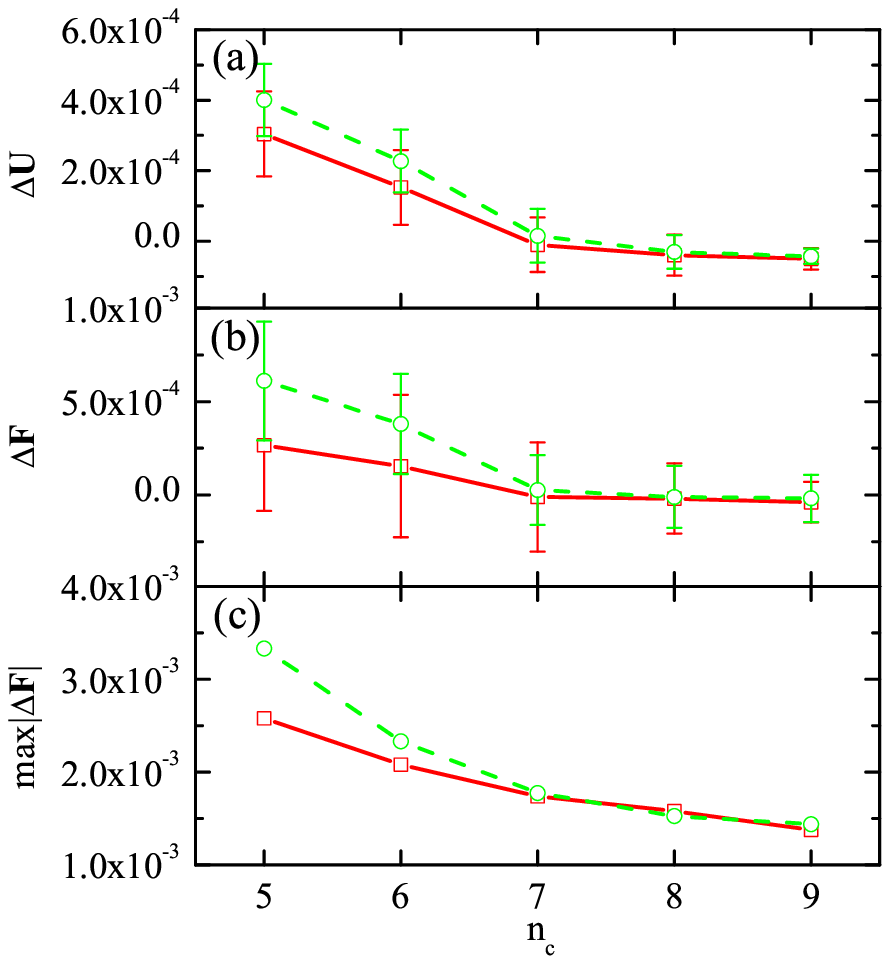}
\caption{Wang \emph{et al.} FIGURE 4}
\end{center}
\end{figure}

\clearpage
\begin{figure}[!ht]
\begin{center}
\includegraphics[angle=0,width=1.0\linewidth]{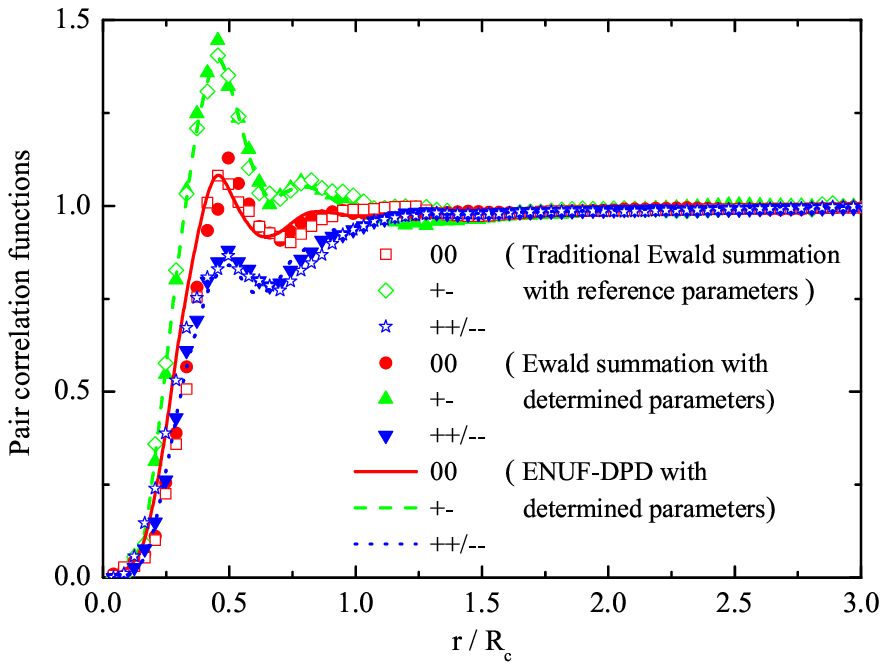}
\caption{Wang \emph{et al.} FIGURE 5}
\end{center}
\end{figure}

\clearpage
\begin{figure}[!ht]
\begin{center}
\includegraphics[angle=0,width=1.0\linewidth]{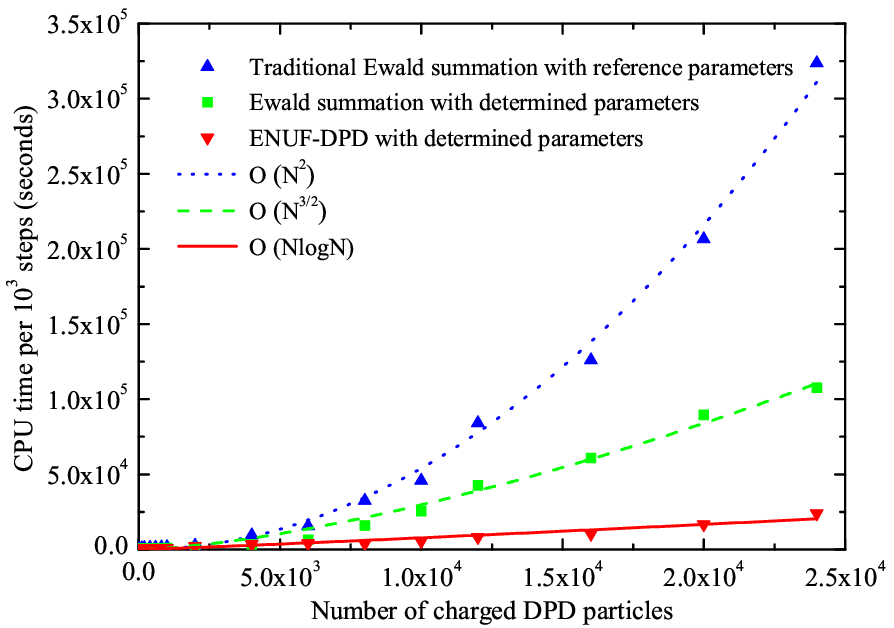}
\caption{Wang \emph{et al.} FIGURE 6}
\end{center}
\end{figure}

\clearpage
\begin{figure}[!ht]
\begin{center}
\includegraphics[angle=0,width=1.0\linewidth]{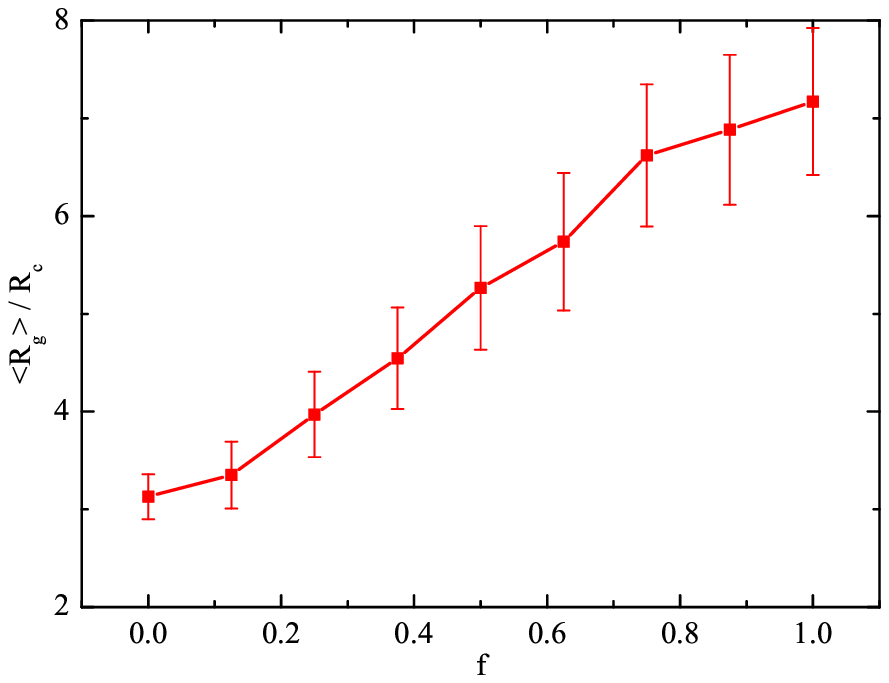}
\caption{Wang \emph{et al.} FIGURE 7}
\end{center}
\end{figure}

\clearpage
\begin{figure}[!ht]
\begin{center}
\includegraphics[angle=0,width=1.0\linewidth]{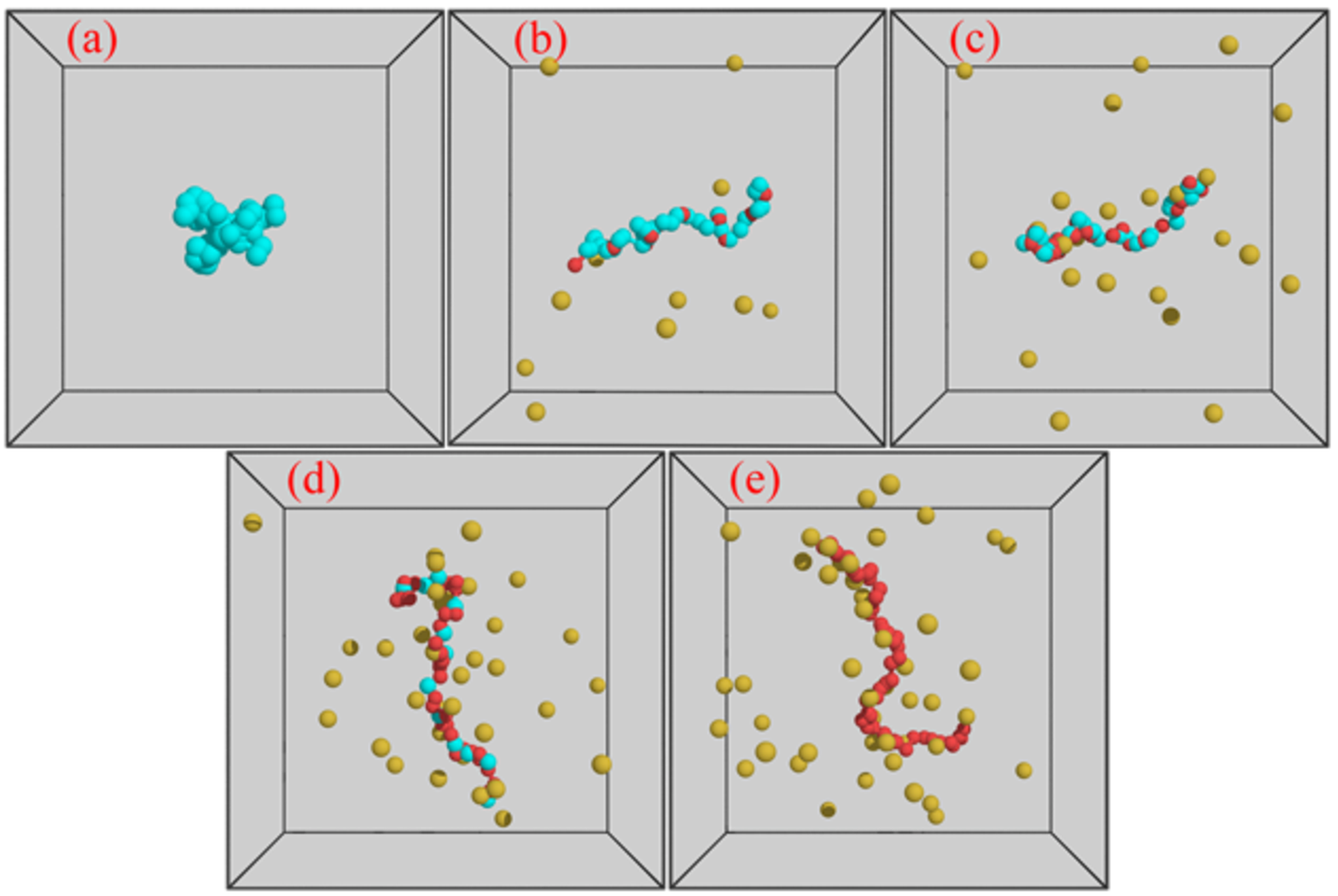}
\caption{Wang \emph{et al.} FIGURE 8}
\end{center}
\end{figure}

\clearpage
\begin{figure}[!ht]
\begin{center}
\includegraphics[angle=0,width=1.0\linewidth]{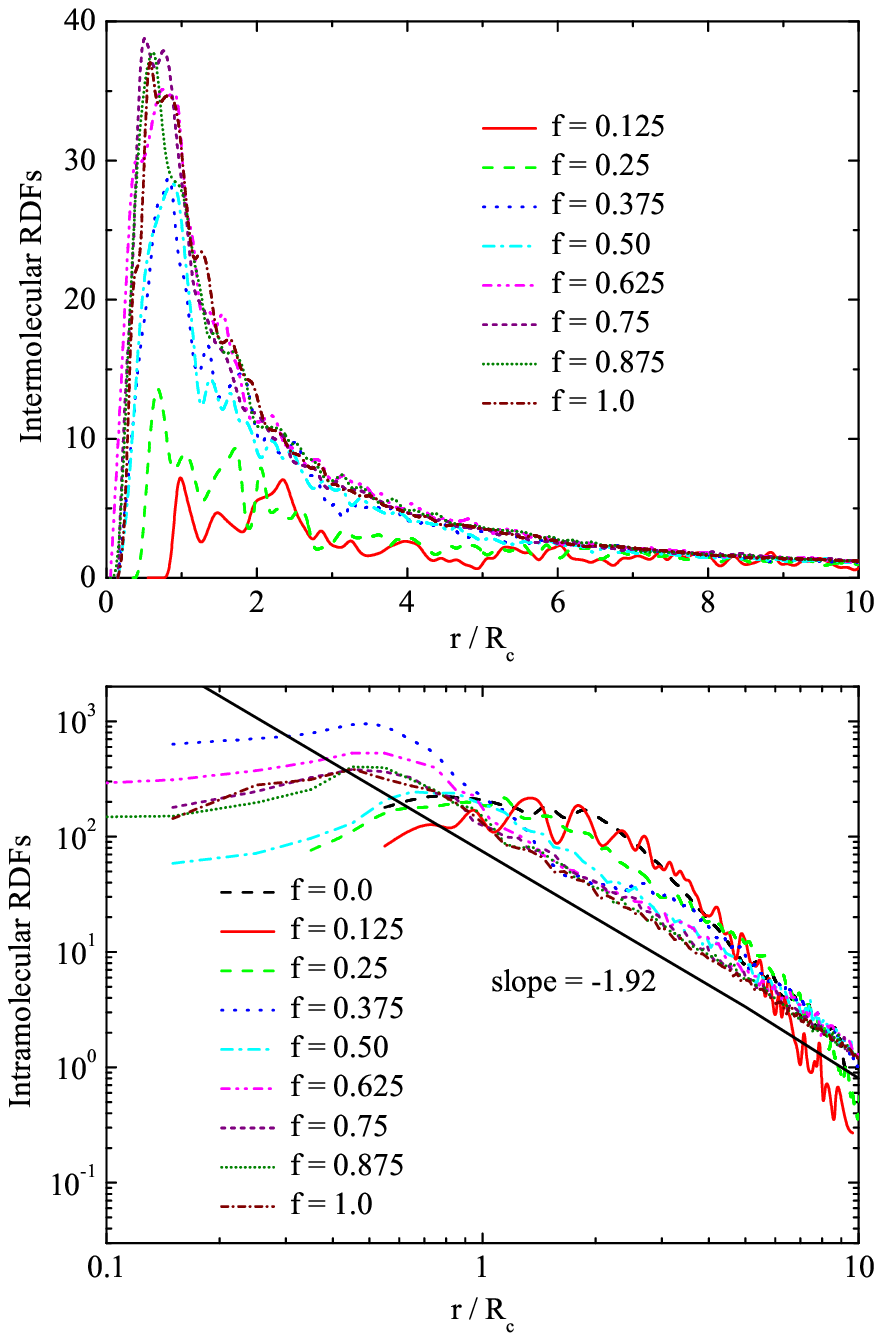}
\caption{Wang \emph{et al.} FIGURE 9}
\end{center}
\end{figure}

\clearpage
\begin{figure}[!ht]
\begin{center}
\includegraphics[angle=0,width=1.0\linewidth]{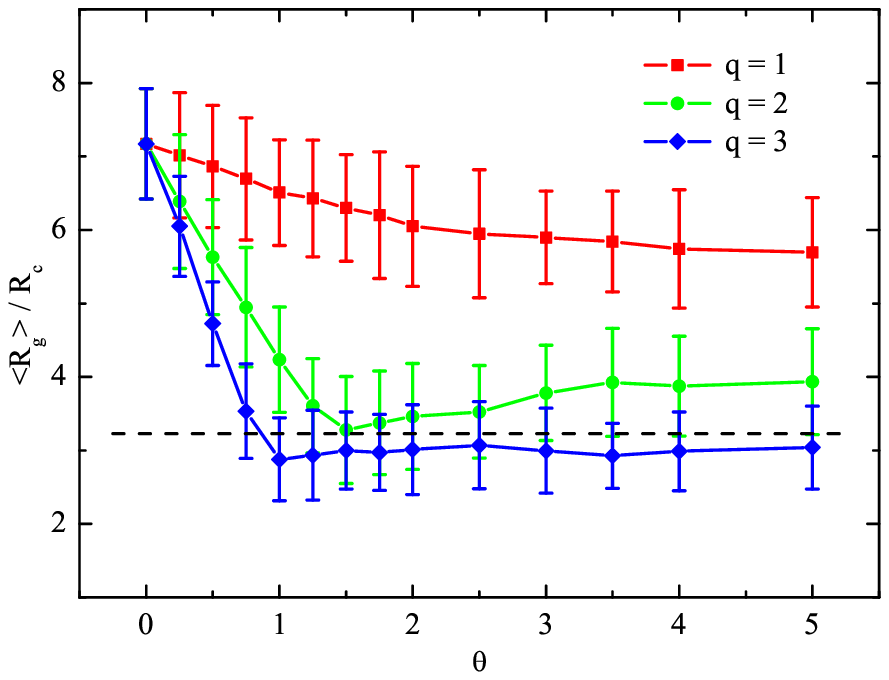}
\caption{Wang \emph{et al.} FIGURE 10}
\end{center}
\end{figure}

\clearpage
\begin{figure}[!ht]
\begin{center}
\includegraphics[angle=0,width=1.0\linewidth]{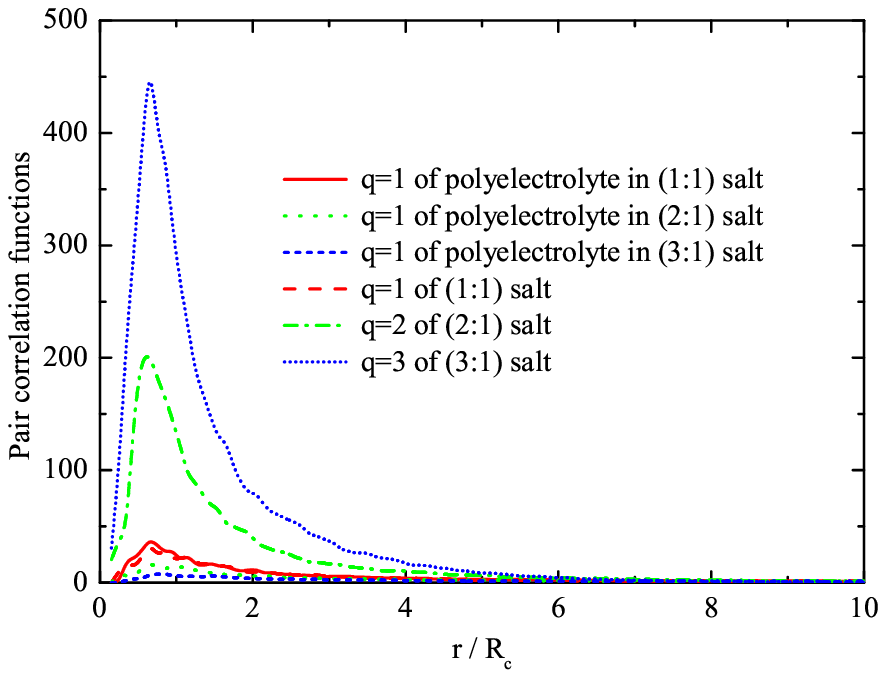}
\caption{Wang \emph{et al.} FIGURE 11}
\end{center}
\end{figure}

\clearpage
\begin{figure}[!ht]
\begin{center}
\includegraphics[angle=0,width=1.0\linewidth]{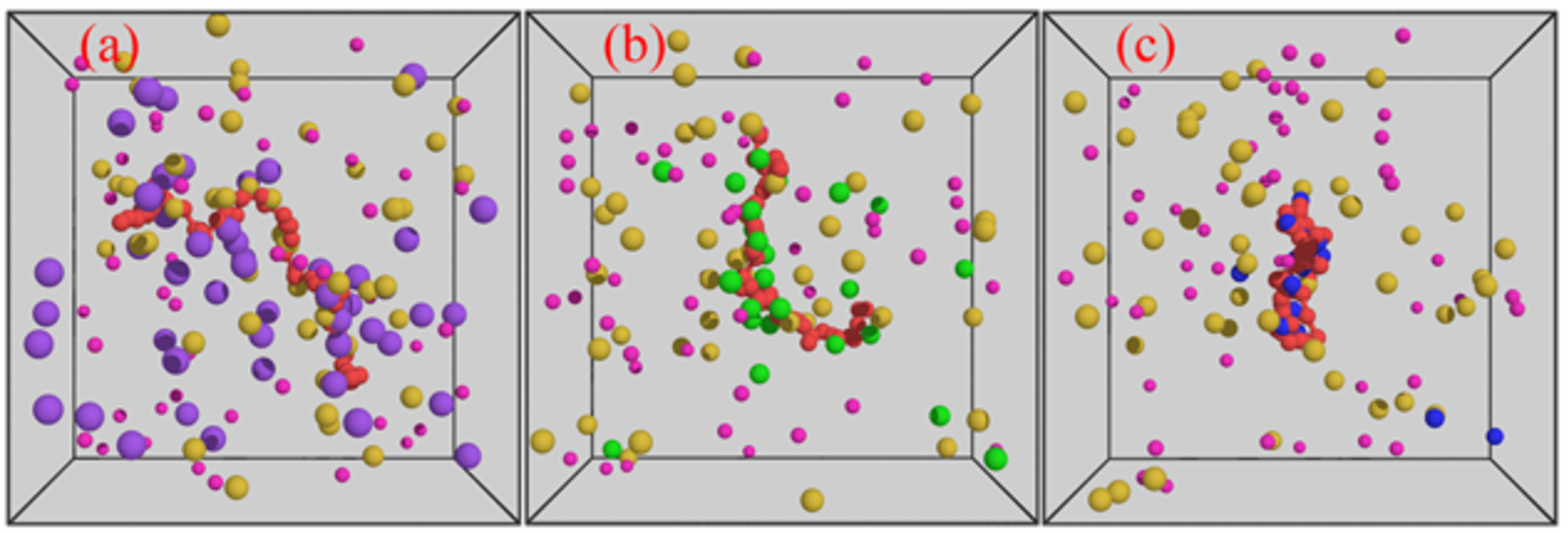}
\caption{Wang \emph{et al.} FIGURE 12}
\end{center}
\end{figure}

\end{document}